\documentclass[%
 linenumbers,
 amsmath,amssymb,
 aps, physrev, reprint,onecolumn
]{article}
\usepackage[numbers, sort&compress]{natbib}
\usepackage{PRIMEarxiv}
\usepackage{subcaption}
\usepackage{placeins}
\usepackage{graphicx}
\usepackage{dcolumn}
\usepackage{bm}
\usepackage{xcolor}
\usepackage[absolute,overlay]{textpos}
\usepackage{amsmath}
\usepackage{dsfont}
\usepackage{authblk}

\newcommand{\eps}{\varepsilon}
\newcommand{\del}{\delta}

\begin{document}
\title{Nested homogenization of xylem-inspired porous fluidic networks}

\author[1]{P.G. Ledda*}
\author[2]{G. Ferrari}
\author[2]{G.A. Zampogna}

\affil[1]{DICAAR, Università degli Studi di Cagliari, Via Marengo 2, 09123 Cagliari, Italy}
\affil[2]{DICCA, Università degli Studi di Genova, Via Montallegro 1, 16145 Genova, Italy}

\begingroup\def\thefootnote{*}\footnotetext{Corresponding author e-mail: piergiuseppe.ledda@unica.it}\endgroup
\maketitle

\begin{abstract}
Fluidic networks may contain hierarchical porous structures in which transport across macroscopic interfaces is controlled by geometrical features at multiple nested length scales. Direct resolution of all these scales rapidly becomes computationally prohibitive for network-scale studies, while reduced descriptions require a priori effective hydraulic properties. Here, we develop a nested homogenization framework for rigid porous interfaces under single-phase viscous flow. The pore-scale structure is first replaced by an effective stress-jump interface law, which is then embedded as an internal interface condition within a second characteristic problem at the intermediate scale. This successive upscaling propagates the pore-scale geometry through effective tensors to obtain a closure for flow across the whole porous structure. The approach is validated against fully resolved simulations at the mesoscopic and macroscopic scales,
for two xylem-inspired channel configurations, and accurately captures pressure drops and flow redistribution while strongly reducing computational cost. The framework provides a tractable bottom-up description of a network composed of nested porous membranes and a basis for extensions to deformable structures and multiphase flows.

\end{abstract}


\section{Introduction}
\begin{figure}
\centering
\includegraphics[width=0.75\textwidth]{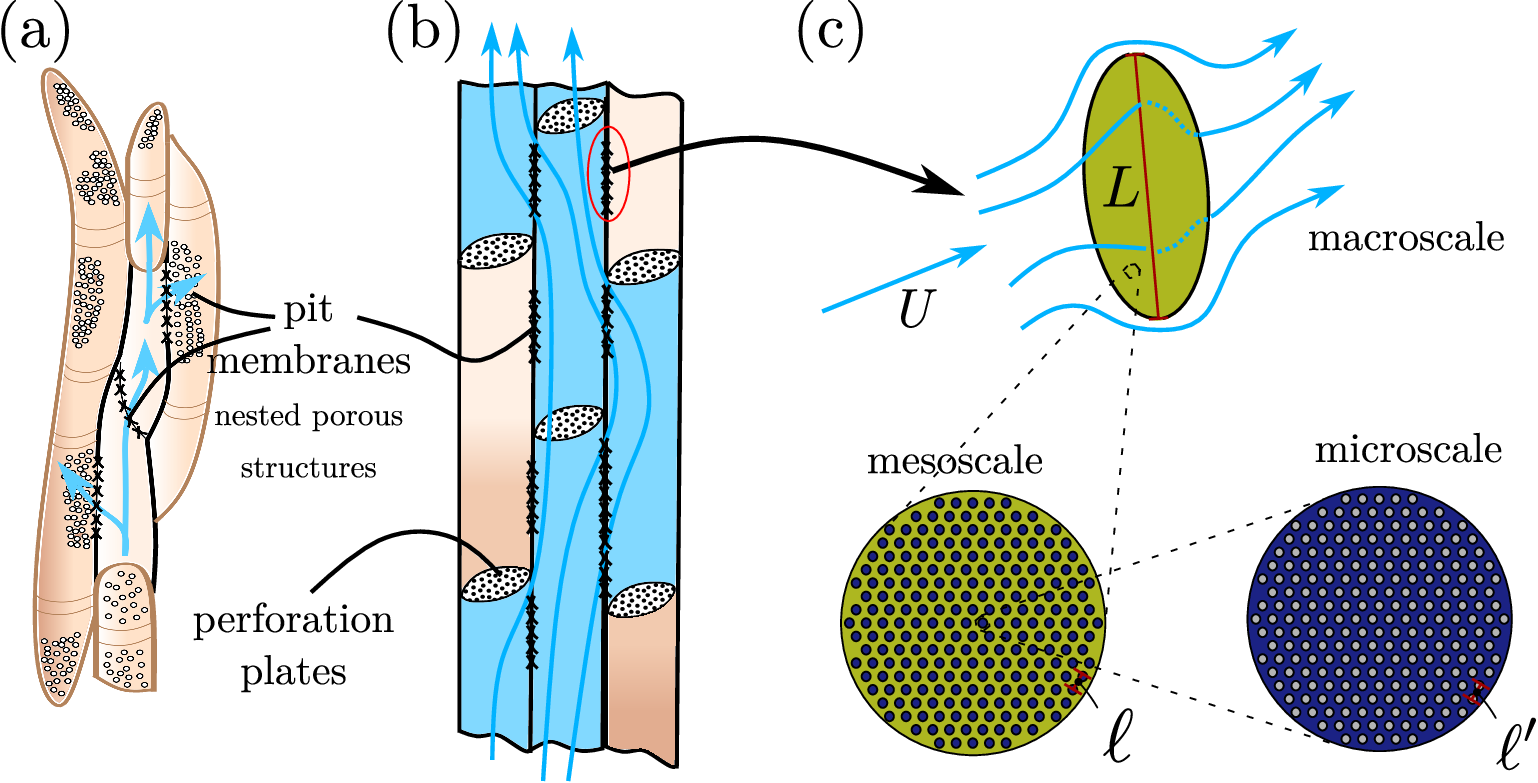}
\caption{Schematics of xylem-inspired nested multiscale architectures considered in this work. Biological inspiration. (a)  Adjacent xylem tracheids exchange fluid through pit membranes. (b) Idealized vessel elements with lateral pit-mediated exchange across functional and non-functional cells.
(c) Nested homogenization setting: a macroscopic thin permeable region of characteristic length \(L\) is replaced by an effective interface law obtained by successively upscaling a mesoscopic pit-scale structure of size \(\ell\) and a microscopic membrane-pore structure of size \(\ell'\).}
\label{fig:nested_architecture_overview}
\end{figure}

Efficient hydraulic control and flow redistribution across scales is a central challenge in engineering and biology, as exemplified by the transport of water and nutrients in plants~\cite{Jensen2016RMP}. Specifically, xylem is a paradigmatic complex hydraulic network comprising tens to thousands of conduits and spanning several length scales~\cite{Choat2008NPhytol,Boutilier2014PLOSOne} called tracheids and vessels~\cite{TyreeZimmermann2002XylemBook,CruiziatCochardAmeglio2002AnnForSci,Choat2008NPhytol}. Tracheids (Fig.~\ref{fig:nested_architecture_overview}a) are individual elongated cells that partially overlap and exchange water laterally across tiny pits (local interruptions of the conduit walls) in their shared walls. Conversely, vessels are shorter cells stacked end-to-end with perforated end walls, whose side walls still contain pits enabling lateral exchange (Fig.~\ref{fig:nested_architecture_overview}b). 
Within each pit, a thin permeable surface, called pit membrane, separates the two conduits. Lateral exchange is therefore controlled by a porous architecture with two length scales, with pits forming the larger-scale openings and the pores of the pit membrane providing a second scale. Pit membranes largely control inter-vessel hydrodynamic resistance and regulate flow redistribution across interconnected conduits, with important consequences for air-embolism propagation~\cite{ZwienieckiHolbrook2000PlantPhysiolPit,Choat2008NPhytol,Jensen2016RMP}. 

Small-scale fluidic circuits and lab-on-a-chip devices are increasingly used in biomedical applications to enable parallelized experiments~\cite{Beebe2002ARBE,Colace2013ARBE,Sackmann2014Nature}.
The combination of local hydraulic elements and large-scale connectivity has motivated a broad class of reduced and biomimetic descriptions of fluidic networks. In engineering settings, passive fluidic networks have been designed to redistribute flow through interconnected channels and porous elements~\cite{Min2020Langmuir,Guo2022MSNE,Vincent2024PRF,Osman2025LabChip}. Related to biological transport, building upon earlier descriptions of exchange across permeable tubule walls by Aldis~\cite{Aldis1988BMB}, network models have connected local inter-conduit transport to the hydraulic response of vascular architectures~\cite{Jensen2016RMP}. Biomimetic experiments have shown how pit-inspired constrictions can control transport at much larger scales. For example, Keiser \emph{et al.} demonstrated stop--go air invasion in compliant microchannel networks~\cite{Keiser2022JFM}, reproducing propagation phenomenology observed in leaves~\cite{Keiser2024JRSI}.
These studies highlight the interplay between the hydraulic response of individual connections and the topology of the surrounding network. Existing network-level models have clarified the importance of vessel connectivity and pathway redundancy in determining hydraulic function in biological cases~\cite{Loepfe2007JTB,Mrad2018PCE,Wason2021PlantPhysiol}. However, these descriptions generally require the hydraulic laws governing pits or inter-conduit transfer to be prescribed as model inputs. Conversely, detailed pore- or pit-scale simulations can quantify the local mechanisms associated with the underlying geometry, but do not directly provide computationally tractable closures for network-scale simulations~\cite{Schulte2012NPhytol,Xu2020SciRep}. This separation between detailed local physics and tractable network descriptions motivates a systematic framework able to propagate the hydraulic response of complex porous structures such as xylem toward the macroscopic network dynamics.

A direct numerical simulation of the flow across multiple nested length scales is extremely demanding and quickly becomes prohibitive for systematic studies~\cite{Xu2020SciRep}. Even when feasible for representative simplified configurations, fully resolved simulations of specific cases hardly yield transferable and scalable design principles. Homogenization theory offers an alternative route by bridging scales through general effective laws and, therefore, naturally addresses the critical computational bottlenecks with no fitting parameters~\cite{Bottaro2019JFMAdjointHomog}.
Specifically, the clear separation between scales, e.g. the vessel size compared to the pit one, enables a macroscopic (homogenized) flow description, derived through a multiscale expansion followed by averaging~\cite{Bottaro2019JFMAdjointHomog,DalyRoose2015PRSATwoFluidHomog,LacisZampognaBagheri2017PRSAPoroelasticBeds}. In this framework, the influence of the small-scale structure is retained explicitly through effective tensors and closure problems posed on a representative periodic element, yielding predictive macroscale equations and interface conditions~\cite{DalyRoose2015PRSATwoFluidHomog,LacisZampognaBagheri2017PRSAPoroelasticBeds}. Homogenization therefore provides a direct link between geometry and permeability, and  can support optimization workflows where the microstructure is tuned to meet prescribed objectives~\cite{Ledda_Boujo_Camarri_Gallaire_Zampogna_2021}.
For thin permeable screens, a description separately applicable both to arrays of pits on the vessel walls and, at a smaller scale, to the array of pores composing the pit membranes themselves, the effective coupling between the two sides of the screen can be expressed via a stress-jump condition. In this approach, the interfacial velocity is proportional to the upstream and downstream stresses through permeability tensors obtained from Stokes-type characteristic problems solved across the repetitive pore geometry~\cite{ZampognaGallaire2020JFMStressJump} (Fig.~\ref{fig:nested_architecture_overview}c). In the simplest settings, this reduces to a Darcy-like relation, i.e., a linear dependence between the normal velocity $U$ and the pressure drop $\Delta p$ across the screen, $U \propto \Delta p$, neglecting viscous normal stresses. Extensions of this framework successfully described passive solute transport and inertial effects~\cite{ZampognaLeddaGallaire2022PoFSoluteFluxJump,WittkowskiPonteLeddaZampogna2024JFMQuasiLinear}, as well as velocity discontinuities and coupled solvent-solute transport ~\cite{ZampognaLeddaWittkowskiGallaire2023JFMJanus,Wittkowski_Ledda_Giordano_Gallaire_Zampogna_2025}. These results suggest that effective interface laws can retain the essential influence of complex porous structures while remaining compatible with network-level simulations.

Previous stress-jump formulations replaced one microstructured membrane by one effective interface. However, a porous vascular network with multiple nested scales requires an approach beyond classical two-scale approaches. A suitable technique is reiterated homogenization, which applies multiscale expansions sequentially at each structural level. This yields a bottom-up macroscopic description in which characteristic problems at smaller scales provide effective parameters and interface conditions for the larger-scale problems~\cite{RamirezTorresEtAl2018IJSS3Scale}. In the present context, this strategy offers a systematic way to propagate the hydraulic effect of pore constrictions up to pit-scale permeabilities and, eventually, to network-scale redistribution laws. 
{The key step is that the pore-scale homogenized interface law becomes an internal boundary condition of a second characteristic problem. The mesoscopic effective response therefore depends on the pore-scale interface tensor and on its interaction with the mesoscopic geometry.}

In this work, we develop a nested homogenization framework for rigid porous membranes, inspired by xylem connectivity, under single-phase viscous-flow conditions. An intervessel connection is idealized as an array of apertures in the conduit wall, each spanned by a nested porous membrane. By applying the upscaling procedure sequentially, the pore-scale response is first condensed into a pit-scale interface law and then propagated to the conduit or network scale~\cite{ZwienieckiHolbrook2000PlantPhysiolPit,KaackEtAl2021NPhytolConstrictions}. 
The framework is validated against fully resolved simulations at the mesoscopic scale and in two distinct macroscopic xylem-inspired configurations. Eventually, the resulting interface law is incorporated into an idealized network with randomly disabled conducting elements to illustrate how the local hydraulic closure can help study network-scale flow redistribution. Although motivated here by xylem, the framework is not tied to a specific geometry and can be adapted to other nested porous structures. 
The paper is organized as follows. Section~II presents the nested homogenization framework and the workflow for obtaining an equivalent macroscale condition. In Section~III, the equivalent interface condition is compared with fully resolved simulations in a two-dimensional analogue of a tracheid connection and of a two-vessel connection. Finally, Section~IV embeds the homogenized interface law in a synthetic xylem-like network and examines how progressive loss of conducting elements affects the emergent hydraulic response.

\section{Reiterated homogenization for nested porous structures}

The single-phase inertialess flow of a viscous Newtonian fluid of viscosity $\mu$, density $\rho$, and kinematic viscosity $\nu=\mu/\rho$ is considered. 
The Stokes equations for velocity and pressure are non-dimensionalized with a characteristic far-to-the-membrane velocity $U$, viscous stress $\mu U /L$, macroscopic membrane size $L$, i.e.,
\begin{equation}
    {\nabla} \cdot {\bm{U}} =0, \quad  {\nabla}{P} =  {\nabla}^2 {\bm{U}},
\end{equation} 
where ${\bm{U}}$ and ${P}$ are the velocity and pressure fields. The stress tensor reads as $\Sigma (\bm{U},P)=-P\bm{I} + (\nabla \bm{U}+\nabla \bm{U}^T)$, where $\bm{I}$ is the identity matrix.
The conduit walls present repetitions of pores, nested at two different scales, whose characteristic length scales are denoted as $\ell$ and $\ell^\prime$.
Since $\ell \ll L$ and $\ell^\prime \ll \ell$, two separation of scales parameters, defined as the ratios between these scales at play, are introduced:
\begin{equation}
\eps=\frac{\ell}{L},\qquad
\del=\frac{\ell'}{\ell},
\end{equation}
so that the nested scale separation is controlled by $\eps$ (macro--meso) and
$\delta$ (meso--micro).
The two scale-separation parameters are treated as independent. 
In the following, we show how two successive asymptotic expansions lead to a systematic framework that replaces the nested porous membrane with a homogeneous, smooth, zero-thickness surface with an effective interface condition. In the outer and inner pure-fluid regions separated by the permeable shell, the incompressible Stokes equations hold.

\subsection*{Mesoscopic flow expansion}

The first step of the procedure, from the macro- to the meso-scale, consists in an asymptotic expansion in $\varepsilon$.
At the mesoscale, the pressure jump across the array is balanced by the local, microscopic, viscous stresses. A fast (mesoscopic) and a slow (macroscopic) variable are introduced, so that $\nabla \rightarrow \nabla_{\bm{x}}+\varepsilon\nabla_{\bm{X}}$, where the subscripts $\bm{x}$ and $\bm{X}$ denote derivation with respect to the fast and slow variables, respectively, in the vicinity of the membrane. The velocity and pressure fields are expanded in series:
\begin{equation}
  \bm{U} = \bm{u}^{(0)} + \eps\,\bm{u}^{(1)} + \eps^2\,\bm{u}^{(2)}+\cdots,\quad
  P      = p^{(0)} + \eps\,p^{(1)} + \eps^2\,p^{(2)}+\cdots.
\end{equation}

In the generic mesoscopic repetitive element (see Fig.~\ref{fig:validation_geometries}a) whose domain is denoted as $\mathcal{F}^{MESO}$ , the flow is governed by the Stokes equations. After the asymptotic expansion, the leading order in $\eps$  problem to be solved reads:
\begin{equation}\label{eq:leading-order}
\begin{aligned}
  -\nabla_{\bm{x}} p^{(0)} + \nabla_{\bm{x}}^2 \bm{u}^{(0)} &= \textbf{0}, \quad \nabla_{\bm{x}} \cdot \bm{u}^{(0)} &= 0,
\end{aligned}
\end{equation}
where boundary and matching conditions with the outer flow must be prescribed. Specifically, no-slip $\bm{u}^{(0)}=\bm{0}$ is imposed at the solid walls $\partial\mathcal{M}$. Conversely, on the upstream (-) and downstream (+) far-field boundaries along the normal-to-the-membrane direction, $\mathcal{B}^\pm$, continuity of velocity and traction with the outer, macroscopic, flow is enforced. {Periodic boundary conditions are imposed at the lateral boundaries of $\mathcal{F}^{MESO}$, along which the geometry is repeated.} Introducing the (slowly varying) macroscopic stress tensor $\Sigma^{\mathrm{out}}_{\bm{X}} (\bm{U},P)$, the far-field traction continuity reads
$\Sigma_{\bm{x}}\!\big(\bm{u}^{(0)},p^{(0)}\big) \bm{n} = \Sigma^{\mathrm{out}}_{\bm{X}} \bm{n}$,
where $\bm{n}$ denotes the outward unit normal and the subscripts indicate whether derivatives in the stress tensor are taken with respect to the fast ($\bm{x}$) or slow ($\bm{X}$) variables. Note that $\Sigma^{\mathrm{out}}_{\bm{X}}$ does not depend on the fast variable and therefore acts as a constant forcing for the mesoscale problem for generic, fixed, $\bm{X}$. Furthermore, the mesoscopic domain contains thin membranes with an additional separation of scales, treated next via a second asymptotic expansion.
  
\subsection*{Microscopic flow through the nested membranes}

\begin{figure}
    \centering
    \includegraphics[width=0.85\linewidth]{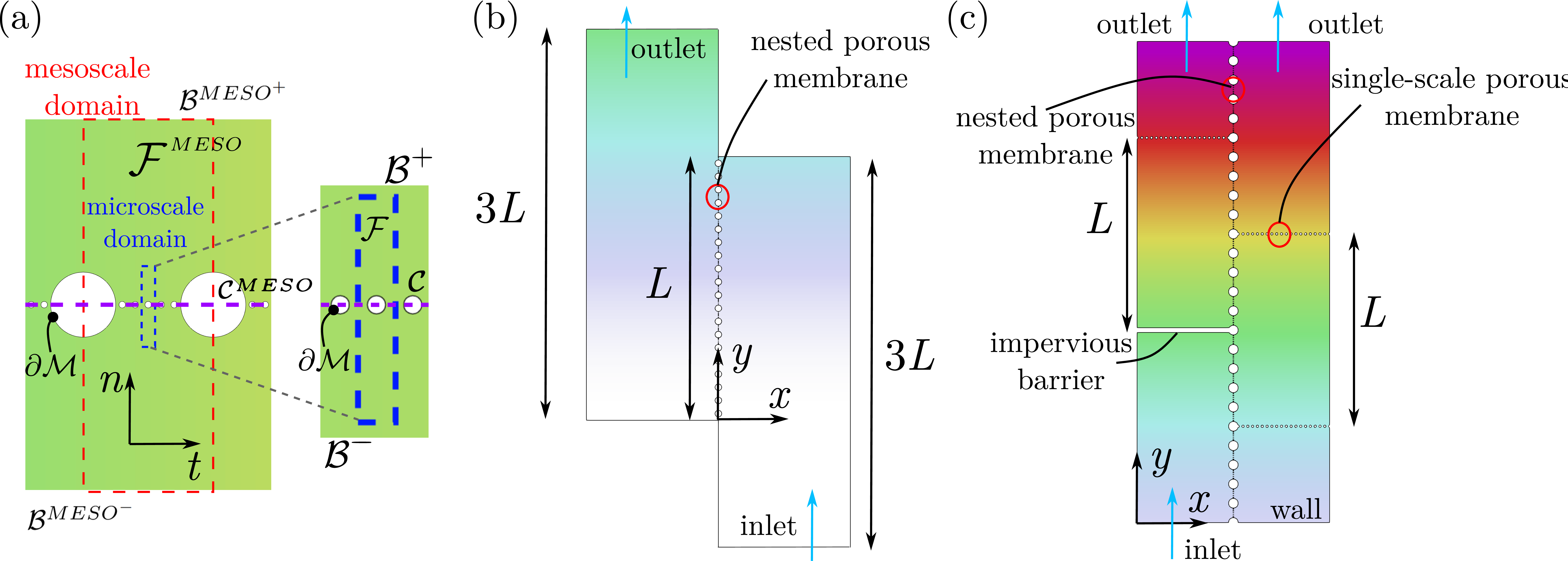}
    \caption{(a) Representative nested membrane element used in the homogenization procedure. The mesoscopic domain contains the pit-scale geometry, while the microscopic domain resolves the smaller pore-scale structure within the membrane. The local coordinates \(\bm n\) and \(\bm t\) denote the directions normal and tangential to the membrane, in this two-dimensional setting, respectively. (b,c) Geometries used for the full-scale validation of the nested homogenized interface model. (b) Tracheid-like connection: two partially overlapping conduits exchange fluid through a vertical nested porous membrane of length \(L\). (c) Vessel-like connection: two adjacent channels present transverse single-scale (perforation plate-like) membranes and communicate through a (pit-like) vertical nested porous membrane, with an impervious barrier forcing flow redistribution between the channels. }
    \label{fig:validation_geometries}
\end{figure}

Our starting point is the mesoscopic leading-order problem. We now ``zoom'' into a generic porous microstructure within the mesoscopic domain and introduce a second expansion in the parameter $\del$ (independent of $\eps$) for the leading order mesoscale velocity $\bm{u}^{(0)}$:
\begin{align}
  \bm{u}^{(0)} &= \bm{u}^{(0,0)} + \del\,\bm{u}^{(0,1)} + \del^2\,\bm{u}^{(0,2)}+\cdots,\\
  p^{(0)}      &= p^{(0,0)} + \del\,p^{(0,1)} + \del^2\,p^{(0,2)}+\cdots,
\end{align}
and $\nabla_{\bm{x}} \rightarrow \nabla_{\bm{x^\prime}}+\delta\nabla_{\bm{x}}$, where the mesoscopic variable now plays the role of ``slow'' variable. 
Keeping the leading order in $\del$ yields the nested leading problem valid in the microscopic repetitive element $\mathcal{F}$ (see Fig.~\ref{fig:validation_geometries}a)
\begin{equation}\label{eq:micro-leading2}
\left\{
\begin{aligned}
  \nabla_{\bm{x^\prime}} p^{(0,0)} &= \nabla_{\bm{x^\prime}}^2 \bm{u}^{(0,0)}\quad \text{in } \mathcal{F},\\
  \nabla_{\bm{x^\prime}} \cdot \bm{u}^{(0,0)} &= 0\quad \text{in } \mathcal{F},\\
  \Sigma_{\bm{x}^\prime}\,\!\big(\bm{u}^{(0,0)},p^{(0,0)}\big) \bm{n}&= \Sigma^{\mathrm{out}}_{\bm{x}}\bm{n}\quad \text{on } {\mathcal{B}^\pm},\\
  \bm{u}^{(0,0)} &= \bm{0}\quad \text{on } \partial\mathcal{M},
\end{aligned}
\right.
\end{equation}
{where the continuity between inner microscopic and outer mesoscopic stresses has been enforced on the upstream ($\mathcal{B}^{-}$) and downstream ($\mathcal{B}^{+}$) sides of $\mathcal{F}$ together with periodicity along the tangent membrane directions and no-slip on $\partial\mathcal{M}$.}
By linearity, the flow at the microscopic scale can be written as~\cite{ZampognaGallaire2020JFMStressJump}
\begin{align}
  \bm{u}^{(0,0)} &=
  \bm{M}^{-}\,\Sigma^{\mathrm{out}-}_{\bm{x}}\bm{n}
  +\bm{M}^{+}\,\Sigma^{\mathrm{out}+}_{\bm{x}}\bm{n},
  \label{eq:u-linear-micro}\\
  p^{(0,0)} &=
  \bm{Q}^{-}\cdot \Sigma^{\mathrm{out}-}_{\bm{x}}\bm{n}
  + \bm{Q}^{+}\cdot\Sigma^{\mathrm{out}+}_{\bm{x}}\bm{n}.
  \label{eq:p-linear}
\end{align}
where $\bm{M}^\pm$ and $\bm{Q}^\pm$ are second-order characteristic tensors and vectors, respectively. Physically, these fields encode the local pore-scale response to unit tractions imposed from either side of the membrane, and therefore determine the effective permeability entering the interface law at the larger scale. Quantities $\bm{M}^\pm$ and $\bm{Q}^\pm$ are obtained by solving the following characteristic problems formally analogous to the Stokes equations:
\begin{equation}\label{eq:micro-problem}
\left\{
\begin{aligned}
  \nabla_{\bm{x^\prime}} \bm{Q}^\pm &= \nabla_{\bm{x^\prime}}^2 \bm{M}^\pm,\quad \text{in $\mathcal{F}$}\\
  \nabla_{\bm{x^\prime}} \cdot \bm{M}^\pm &= 0,\quad \text{in $\mathcal{F}$}\\
  \Sigma_{\bm{x}^\prime}\,\!\big(\bm{M}^\pm,\bm{Q}^\pm\big) \bm{n} &= \bm{I}\bm{n} \quad \text{on } \mathcal{B}^\pm,\\
  \Sigma_{\bm{x}^\prime}\,\!\big(\bm{M}^\pm,\bm{Q}^\pm\big) \bm{n} &= \textbf{0} \quad \text{on } \mathcal{B}^\mp,\\
  \bm{M}^\pm &= \bm{0} \quad \text{on } \partial\mathcal{M}.
\end{aligned}
\right.
\end{equation}



{The microscopic average of a generic vector $\bm{f}$ at the membrane centerline, or center-surface in the three-dimensional case, $\mathcal{C}$ on the} microscopic domain is defined as
\begin{equation} \langle \bm{f} \rangle = \frac{1}{\mathcal{C}} \int_{\mathcal{C}} \bm{f} \mathrm{d}x^\prime_s \mathrm{d}x^\prime_t, \end{equation}
where $x^\prime_s$ and $x^\prime_t$ denote the tangential directions. 
After averaging, the leading-order velocity field satisfies
\begin{align}
  \bm{u}^{(0)} &= \langle \bm{u}^{(0,0)} \rangle = \langle\bm{M}^-\rangle\,\Sigma^{\mathrm{out}-}_{\bm{x}}(\bm{u}^{(0)},p^{(0)})\bm{n}
  + \langle\bm{M}^+\rangle\,\Sigma^{\mathrm{out}+}_{\bm{x}}(\bm{u}^{(0)},p^{(0)})\bm{n}, 
   \label{eq:micro-membrane}
\end{align}
where $\langle \bm{M}^\pm \rangle$ are the upstream and downstream microscopic  permeability (also called \emph{Navier} \cite{ZampognaGallaire2020JFMStressJump}) tensors. 
The interface condition can be applied on the porous membranes that are located within the mesoscopic periodic cell. In other words, the microstructured permeable screen (in the case of xylem, the porous pit membrane) can be replaced by an equivalent, zero-thickness surface on which the velocity-stress condition is imposed~\cite{ZampognaGallaire2020JFMStressJump}, while the pressure condition simply states that the interfacial pressure equals the mean of the values measured on each side of the membrane. Note that, for membranes symmetric with respect to the center surface, $\langle\bm{M}^+\rangle = -\langle\bm{M}^-\rangle=\langle\bm{M}\rangle$, yielding a Darcy-like relation, i.e., $ \bm{u}^{(0)} = -\langle\bm{M}\rangle\,\left(\Sigma^{\mathrm{out}-}_{\bm{x}}\bm{n} - \Sigma^{\mathrm{out}+}_{\bm{x}}\bm{n}\right)$. We will now recast this condition on the leading order mesoscale velocity field to formulate an interface condition of the mesoscale characteristic problems.

\subsection*{Macroscopic interface model}

The mesoscale problem is now reformulated in light of the new interface condition at the nested porous surfaces.
Each inner membrane is replaced by a mesoscopic interface, here denoted by $\Gamma^{\mathrm{MESO}}$. When rescaled using mesoscopic reference scales, the interface condition Eq.~\eqref{eq:micro-membrane} reads
\begin{equation}
\bm{u}^{(0)} =
  \delta \langle{\bm{M}^-}\rangle \,\Sigma_{\bm{x}}^-\!(\bm{u}^{(0)},p^{(0)})\bm{n}
  \;+\;
  \delta \langle{\bm{M}^+}\rangle \,\Sigma_{\bm{x}}^+\!\big(\bm{u}^{(0)},p^{(0)})\bm{n},
  \label{eq:membcondpermeso}
\end{equation}
We again exploit linearity, in this case of the mesoscale problem, and express the leading order mesoscopic velocity as
\begin{align}
  \bm{u}^{(0)} &=
  \bm{N}^{-}\,\Sigma^{\mathrm{out}-}_{\bm{X}}\bm{n}
  + \bm{N}^{+}\,\Sigma^{\mathrm{out}+}_{\bm{X}}\bm{n},
  \label{eq:u-linear-meso}\\
  p^{(0)} &=
  \bm{R}^{-}\cdot\Sigma^{\mathrm{out}-}_{\bm{X}}\bm{n}
  + \bm{R}^{+}\cdot \Sigma^{\mathrm{out}+}_{\bm{X}}\bm{n}.
\end{align}
Upon substitution of Eq.~\eqref{eq:u-linear-meso} into Eq.~\eqref{eq:membcondpermeso}, the membrane interface condition can be rewritten in terms of the characteristic tensors, owing to linearity of the stress operator:
{\begin{equation}
\bm{N}^\pm=\delta \Big[ \langle \bm{M}^- \rangle \Sigma_{\bm{x}}^-(\bm{N}^{\pm},\bm{R}^{\pm})\bm{n}+\langle \bm{M}^+\rangle  \Sigma_{\bm{x}}^+(\bm{N}^{\pm},\bm{R}^{\pm})\bm{n}\Big]\,\text{on $\Gamma^{MESO}$} 
\label{eq:membcondpermesoN}
\end{equation}}
\noindent which means that $\bm{N}^\pm$ now depend not only on the mesoscopic geometry, but also on the permeability properties of the inner membranes, and satisfy an associated set of characteristic problems. In practice, these fields quantify how a macroscopic traction applied on either side of the array propagates to the microscale once the nested membrane is replaced by its effective interface law. The mesoscopic tensors $\bm{N}^\pm$ and vectors $\bm{R}^{\pm}$ satisfy 
\begin{equation}
    \begin{cases}
    \nabla_{\bm{x}}\bm{R}^{\pm}=\nabla_{\bm{x}}^2\bm{N}^{\pm} \quad \text{in $\mathcal{F}^{MESO}$}\\
    \nabla_{\bm{x}}\cdot\bm{N}^{\pm}=0 \qquad \text{in $\mathcal{F}^{MESO}$}\\
    \Sigma_{\bm{x}}\big(\bm{N}^\pm,\bm{R}^\pm)\bm{n}=\bm{I}\bm{n} \quad\text{on $\mathcal{B}^{MESO^\pm}$}\\
    \Sigma_{\bm{x}}\big(\bm{N}^\pm,\bm{R}^\pm)\bm{n}=\textbf{0} \quad\text{on $\mathcal{B}^{MESO^\mp}$}\\
    \bm{N}^{\pm}=0\qquad \text{on $\partial\mathcal{M}$,
    }
    \end{cases}
\end{equation}
obtained by substituting Eq.~\eqref{eq:u-linear-meso} into Eq.~\eqref{eq:leading-order}, coupled with Eq.~\eqref{eq:membcondpermesoN} on the homogenized internal membranes.
A second, mesoscopic average at the membrane centerline (or center-surface) of the mesoscopic cell $\mathcal{C}^{MESO}$ forming the pit array is defined:
\begin{equation}
    \lbrace \bm{f} \rbrace = \frac{1}{\mathcal{C}^{MESO}} \int_{\mathcal{C}^{MESO}} \bm{f} \mathrm{d}x_t \mathrm{d}x_s.
\end{equation}
The resulting macroscopic condition thus reads:
\begin{align}
  \bm{U} &= \lbrace \bm{u}^{(0)} \rbrace = \lbrace\bm{N}^{-}\rbrace\,\Sigma^{-}_{\bm{X}}(\bm{U},P)\bm{n}
  + \lbrace\bm{N}^{+}\rbrace\,\Sigma^{+}_{\bm{X}}(\bm{U},P)\bm{n}, 
\end{align}
where $\lbrace\bm{N}^{\pm}\rbrace$ depend on the mesoscopic geometry and on the microscopic permeability properties of the nested membranes.
This interface condition replaces the nested porous structure (in our example, the pit array on the conduit walls), which is thus represented as a surface of negligible thickness (Fig.~\ref{fig:nested_architecture_overview}c). The model retains the hydraulic contribution of the unresolved pores through the tensors $\lbrace\bm{N}^{\pm}\rbrace$, enabling large-scale computations without explicitly meshing the microstructure. Once the membrane condition is nondimensionalized with the macroscopic scale, the permeability tensor entries read $\mathcal{N}_{ij}^\pm= \varepsilon \lbrace{N}_{ij}^{\pm}\rbrace$.
The numerical implementation is carried out with the finite-element solver COMSOL Multiphysics. The nondimensional Stokes equations are implemented in weak form using the \emph{Weak Form PDE} interface. 
The membrane condition is imposed on each membrane line or surface ($\Gamma_m$) via a one-domain approach, as follows. By introducing the local macroscopic normal and tangential to the membrane velocities
$
U_n=\bm{U}\cdot\mathbf{n},
\,
U_t=\bm{U}\cdot\mathbf{t},
$
where $\mathbf{n}$ and $\mathbf{t}$ are the corresponding normal and tangential unit vectors, the membrane law is introduced through the following weak contribution, valid for a symmetric membrane ($\mathcal{N}_{nt}=\mathcal{N}_{tn}=0$ in the $t-n$ reference frame \cite{ZampognaGallaire2020JFMStressJump}),
\begin{equation}
\int_{\Gamma_m}
\left(
\frac{U_n w_n}{\mathcal{N}_{nn}}
+
\frac{U_t w_t}{\mathcal{N}_{tt}}
\right)\,\mathrm{d}\Gamma,
\label{eq:membrane_weak_network}
\end{equation}
where $w_n=\mathbf{w}\cdot\mathbf{n}$, $w_t=\mathbf{w}\cdot\mathbf{t}$, and $\mathbf{w}$ is the test velocity of the weak formulation. Pressure is discretized with discontinuous first-order Lagrange shape functions in order to allow the pressure jump at the membrane, while the velocity (third-order Lagrange shape functions to ensure stability) remains continuous.

In the following, we compare the reduced model with fully resolved simulations of simplified fluidic networks with nested membranes, inspired by tracheid and vessel connections in xylem. This comparison quantifies both the accuracy of the predicted global pressure drop and the ability of the homogenized interface law to reproduce flow redistribution.

\section{Model comparison with full-scale simulations}

Two representative idealized geometries that capture distinct modes of hydraulic connectivity in xylem are considered, see Fig.~\ref{fig:validation_geometries}(b,c). 
The first configuration is a two-dimensional geometry inspired by tracheid-based transport. Two planar conduits partially overlap along their length and exchange fluid laterally through a localized permeable region in their side walls. This region contains a finite array of discrete pits, each spanned by a thin porous membrane.
The second configuration is a two-dimensional geometry representative of vessel-based transport. Specifically, a minimal model of two neighboring vessels is considered. The vessels communicate through pits distributed over their facing walls. 

For the tracheid-inspired connection (panel {b}), two channels of nondimensional width $R=0.5$ are connected on one side by a membrane of unit nondimensional length. The first channel, where the inlet is prescribed, terminates abruptly at the membrane edge. As a result, the flow is forced to cross the nested membrane structure composed of an array of cylindrical inclusions with two distinct characteristic sizes.
Each periodic membrane element, with $\varepsilon = 0.05$ (ratio between mesoscopic and macroscopic scales), contains an internal permeable layer laterally confined by two circular inclusions of radius $r_{\mathrm{meso}} = 0.1$ made nondimensional with $\ell$. This permeable layer contains microscopic inclusions with radius $r_{\mathrm{micro}} = 0.1$, made nondimensional with $\ell^\prime$, and $\delta = 0.05$ (ratio between microscopic and mesoscopic scales), see Fig.~\ref{fig:validation_geometries}(a) for a sketch.
An inlet condition is imposed at the bottom boundary of the left vessel, while the other vessel presents a no-slip condition. The outlet boundary condition is imposed at the upper boundary, downstream of the membrane, at a distance $L_{\mathrm{outlet}} = 2$ from the closer membrane edge.
Fig.~\ref{fig:validation_geometries}({c}) presents the vessel-like connection across two channels. A similar nested structure, with $\varepsilon=0.1$ in this case and $\delta=0.05$  separates two channels of height $2.5$ and width 0.5. These channels present also transverse porous barriers composed of simple cylindrical inclusions of nondimensional radius $r_{transv}=0.{25}$, nondimensionalized with $\ell_{transv}$, and $\varepsilon_{transv}=\frac{\ell_{transv}}{L}
={0.025}$. This distribution of nested membranes and porous barriers results in several cells effectively reproducing a staggered vessel-like connection of two channels. Furthermore, the left channel contains an obstruction that forces the flow to pass through the right channel.
An inlet condition is imposed at the bottom boundary of the left vessel, while the other vessel presents a no-slip condition. At the top, outflow conditions are imposed. 
To complete the numerical implementation description, no-slip (Dirichlet) conditions are prescribed on the solid walls, while a zero-stress condition is imposed at the outlet. At the inlet, the uniform normal reference (along the vertical $y$ direction) velocity $V$ is prescribed. Once nondimensionalized, it reads simply $V=1$ and quickly adjusts downstream to the presence of walls and the lateral porous structures.

\begin{figure}[t!]
\noindent\makebox[\linewidth]{%
  \makebox[0.25\linewidth][s]{\raggedleft (a)}%
  \makebox[0.25\linewidth][s]{\raggedleft(b)}%
  \makebox[0.25\linewidth][s]{\raggedleft(c)}%
  \makebox[0.25\linewidth][s]{\raggedleft(d)}%
}
    \includegraphics[height=5cm]{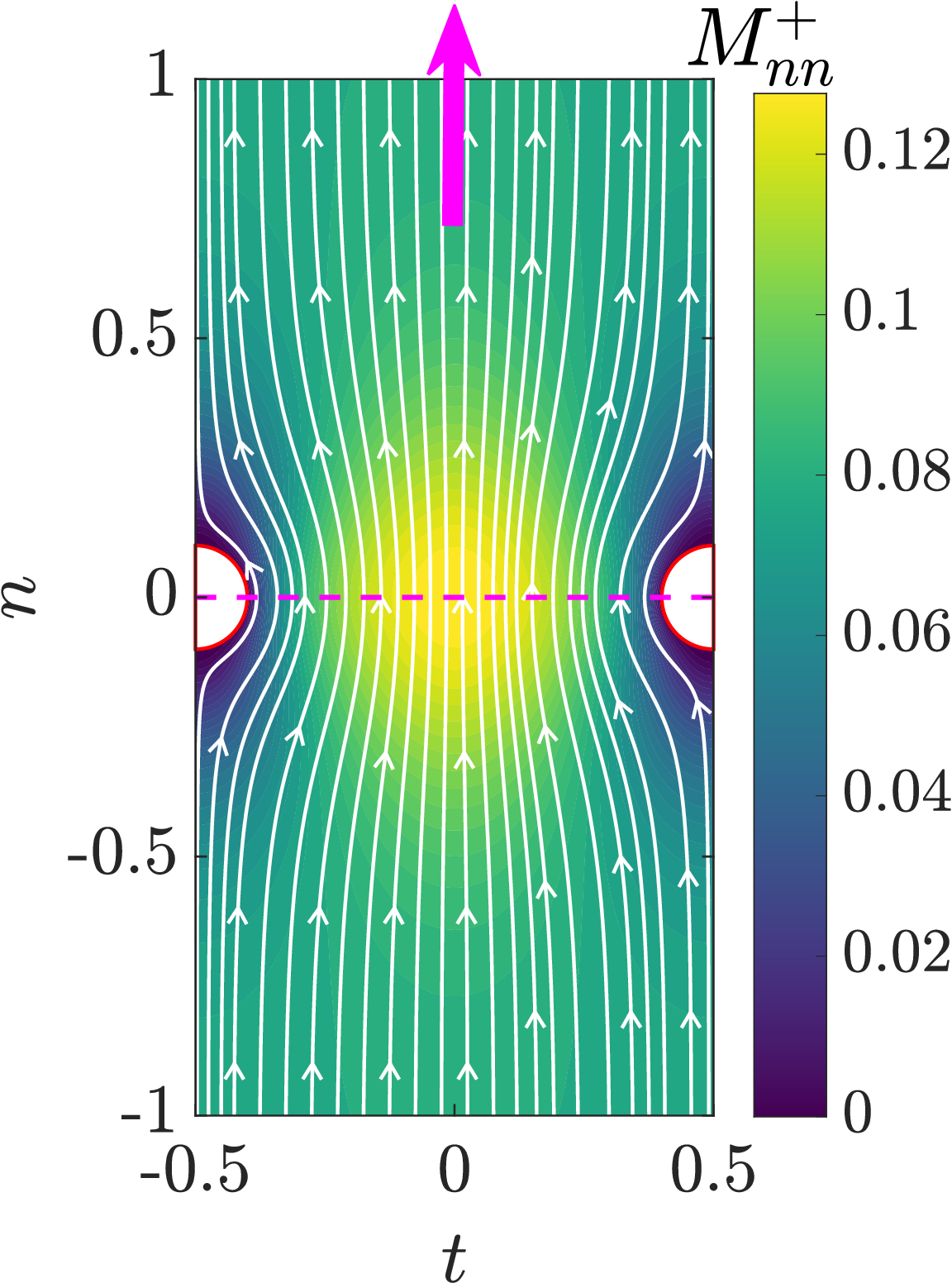}\quad \
 \includegraphics[height=5cm]{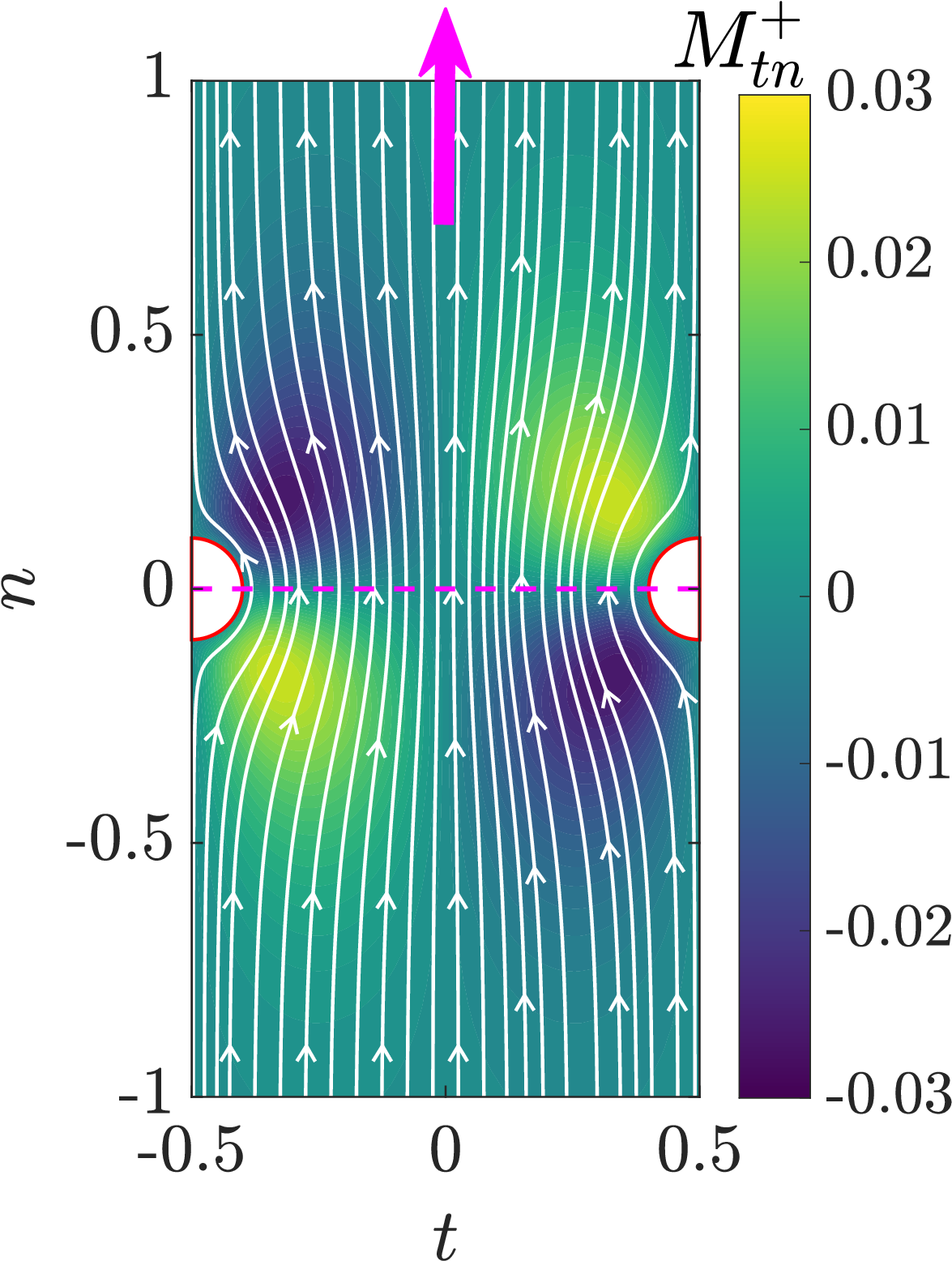}\qquad  
  \includegraphics[height=5cm]{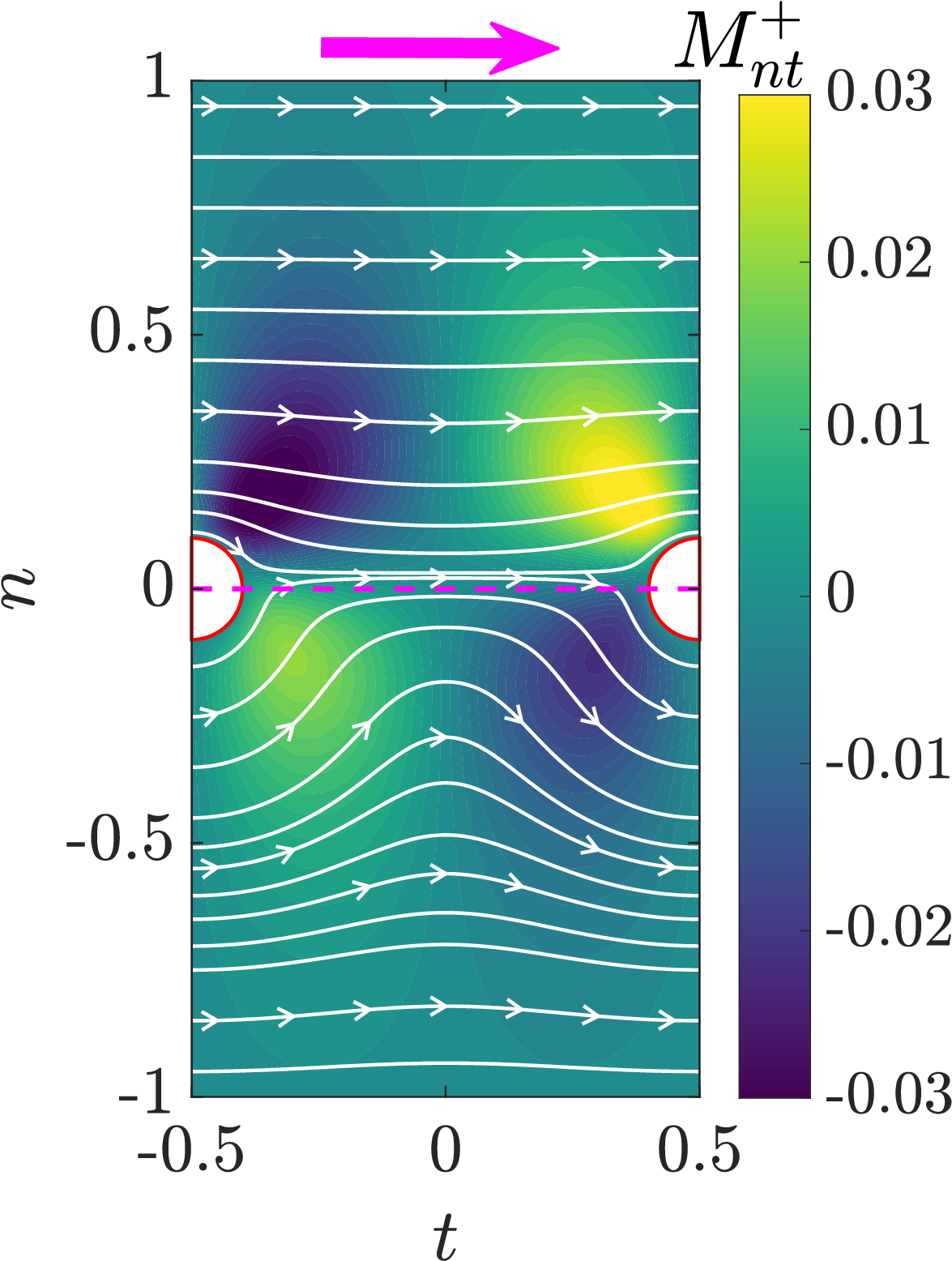}\quad  
 \includegraphics[height=5cm]{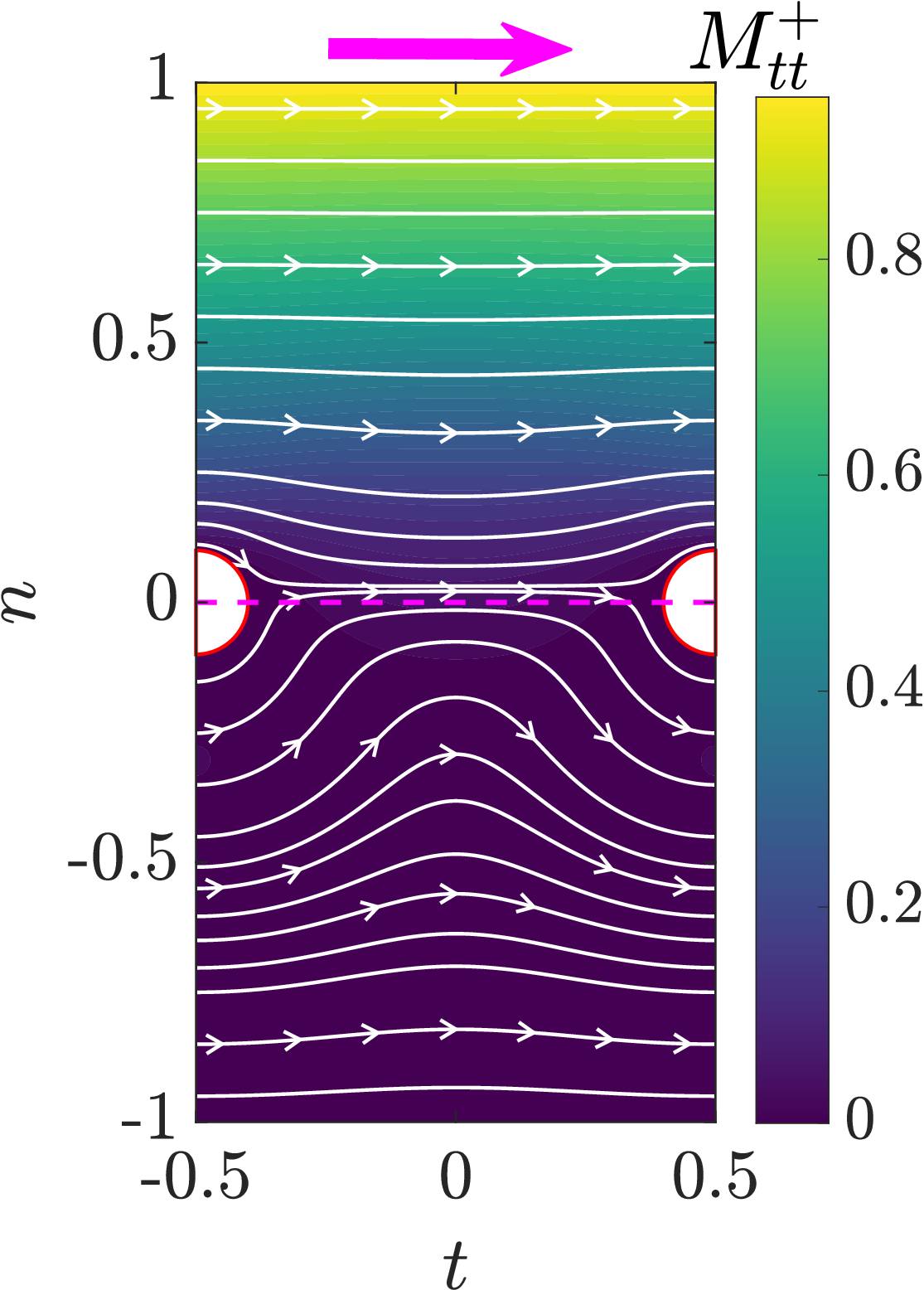}
    \caption{Microscopic characteristic fields for the microscale membrane problem.
Colormaps of the characteristic velocity field components associated with a unit traction imposed on the upper side of the microscopic cell associated with the imposed unit forcing (magenta arrows):
(a) $M_{nn}^{+}$, (b) $M_{tn}^{+}$, (c) $M_{nt}^{+}$, and (d) $M_{tt}^{+}$. 
The coordinates $n$ and $t$ denote the directions normal and tangential to the membrane, respectively. White streamlines show the characteristic velocity field vectors.}
    \label{fig:microscale_characteristic_fields}
\end{figure}

\begin{figure}[t!]
\noindent\makebox[\linewidth]{%
  \makebox[0.25\linewidth][s]{\raggedleft (a)}%
  \makebox[0.25\linewidth][s]{\raggedleft(b)}%
  \makebox[0.25\linewidth][s]{\raggedleft(c)}%
  \makebox[0.25\linewidth][s]{\raggedleft(d)}%
}
\includegraphics[height=5cm]{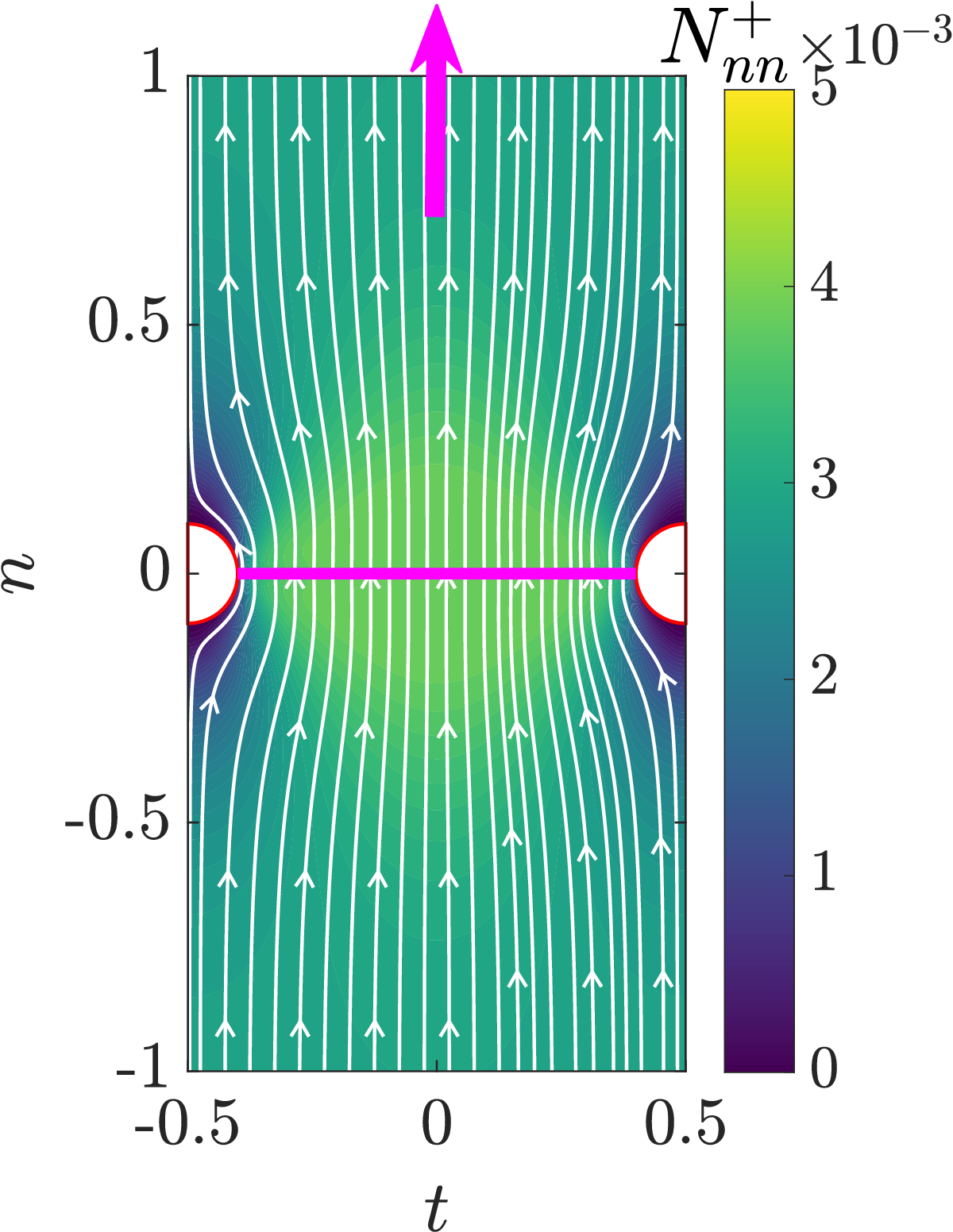}
\includegraphics[height=5cm]{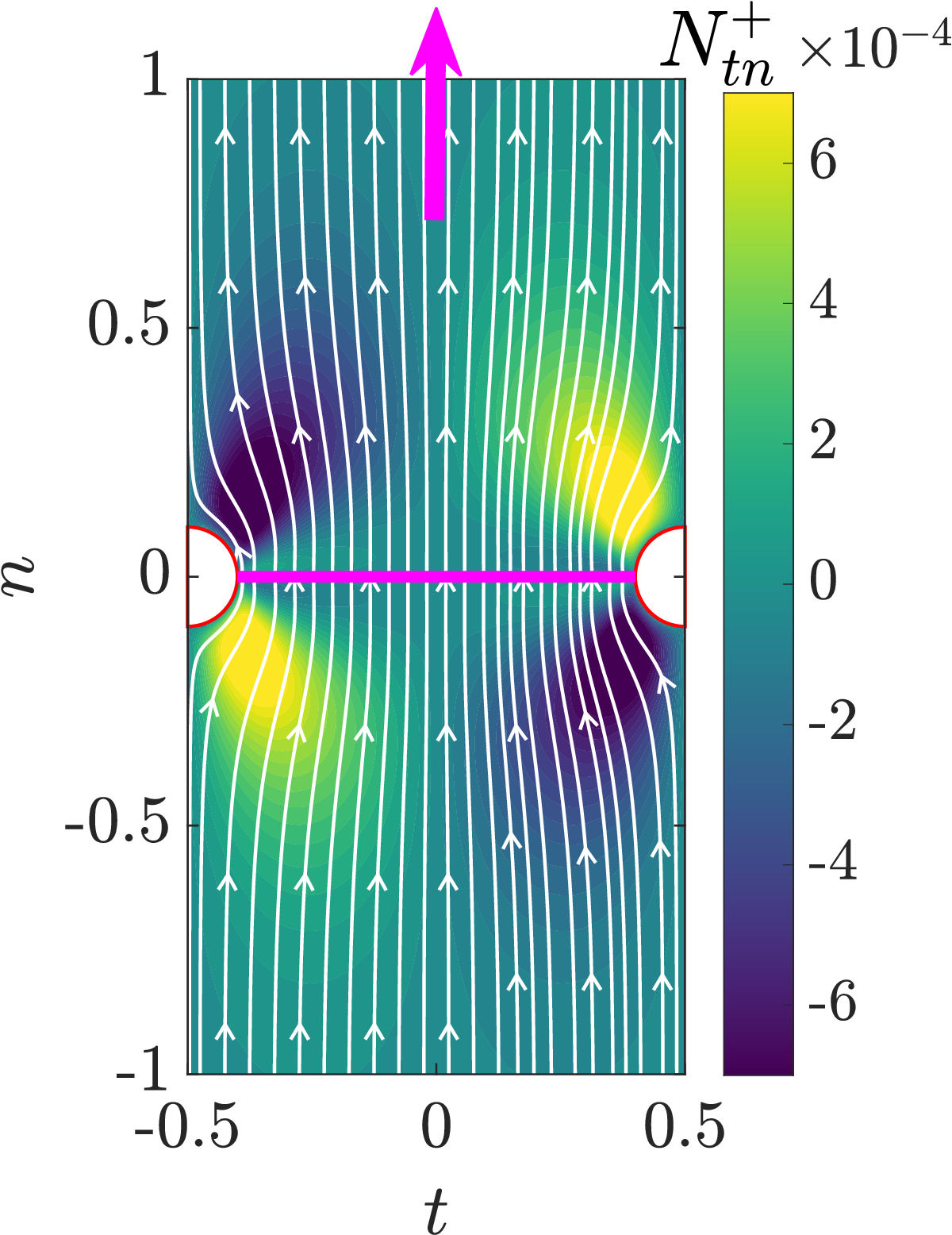}
\includegraphics[height=5cm]{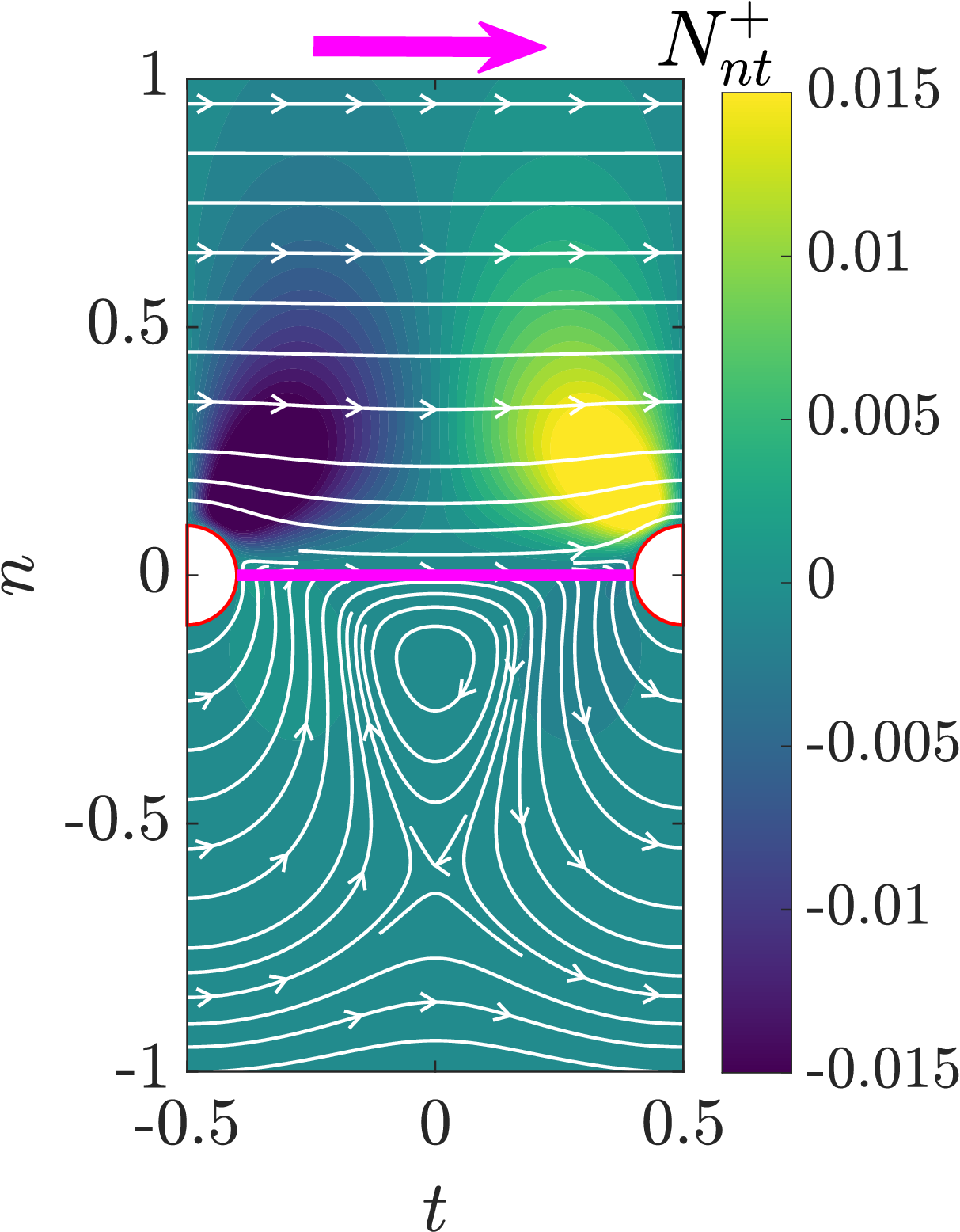}
\includegraphics[height=5cm]{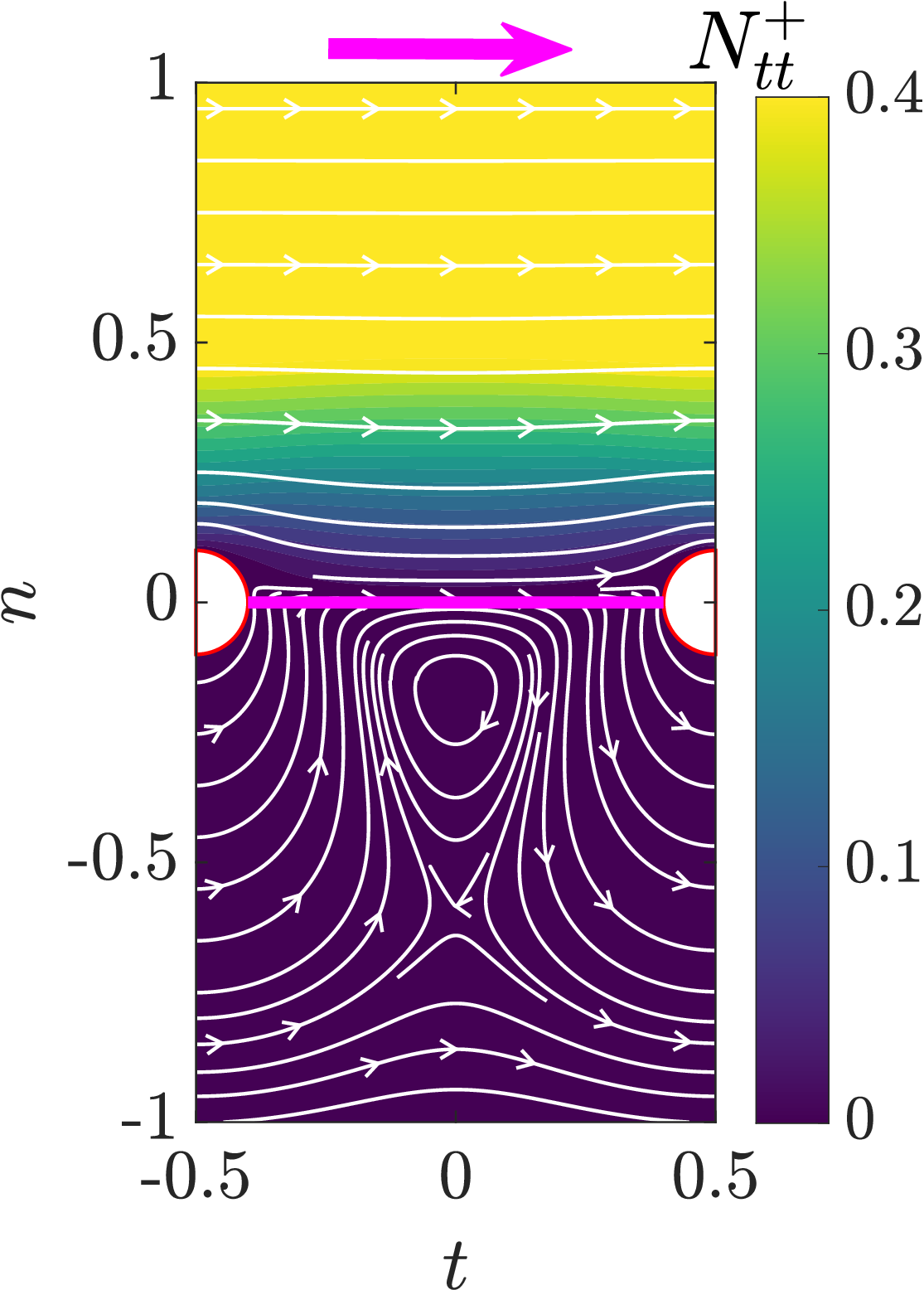}
\noindent\makebox[\linewidth]{%
  \makebox[0.25\linewidth][s]{\raggedleft (e)}%
  \makebox[0.25\linewidth][s]{\raggedleft(f)}%
  \makebox[0.25\linewidth][s]{\raggedleft(g)}%
  \makebox[0.25\linewidth][s]{\raggedleft(h)}%
}
 \includegraphics[height=5cm]{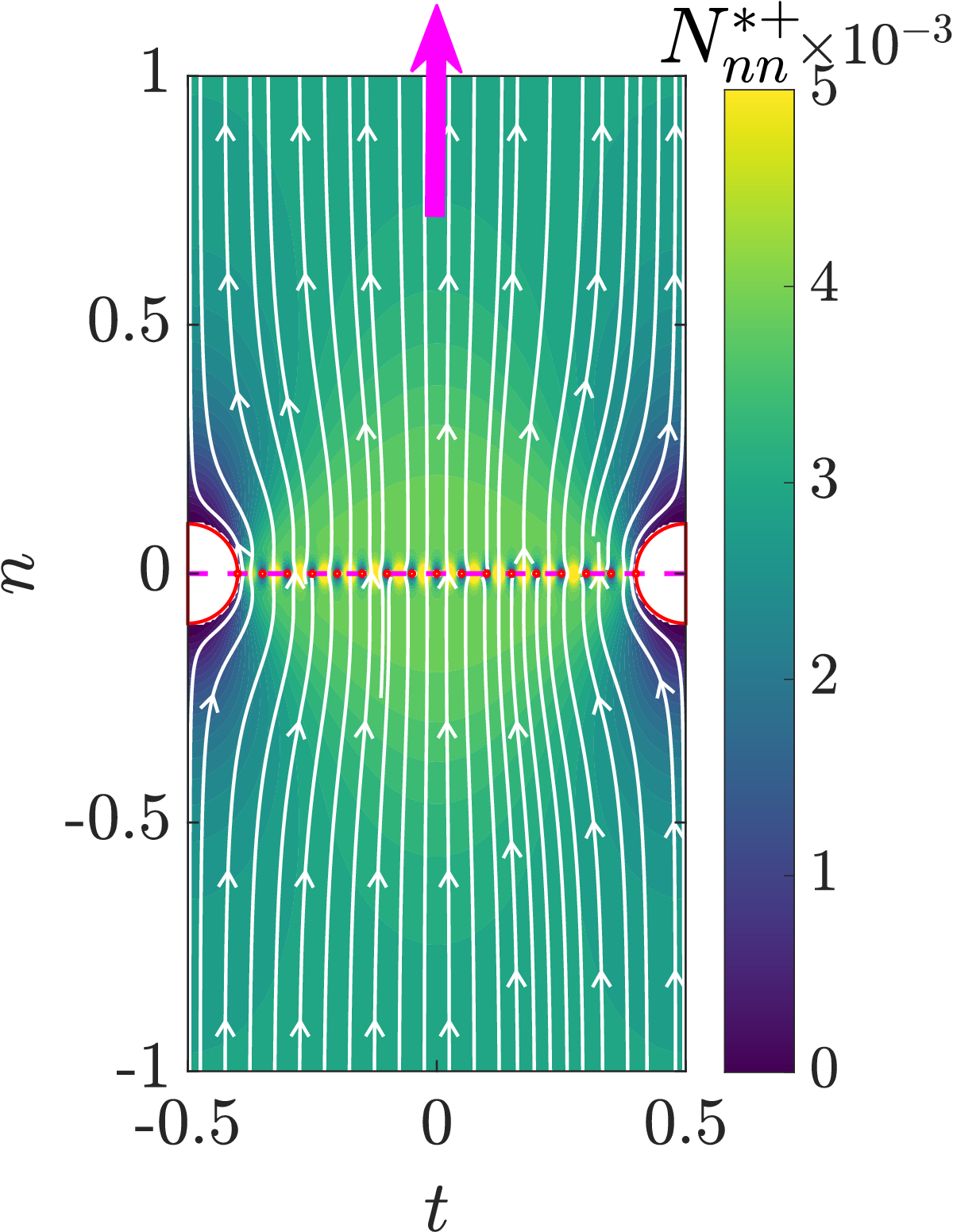}\quad
\includegraphics[height=5cm]{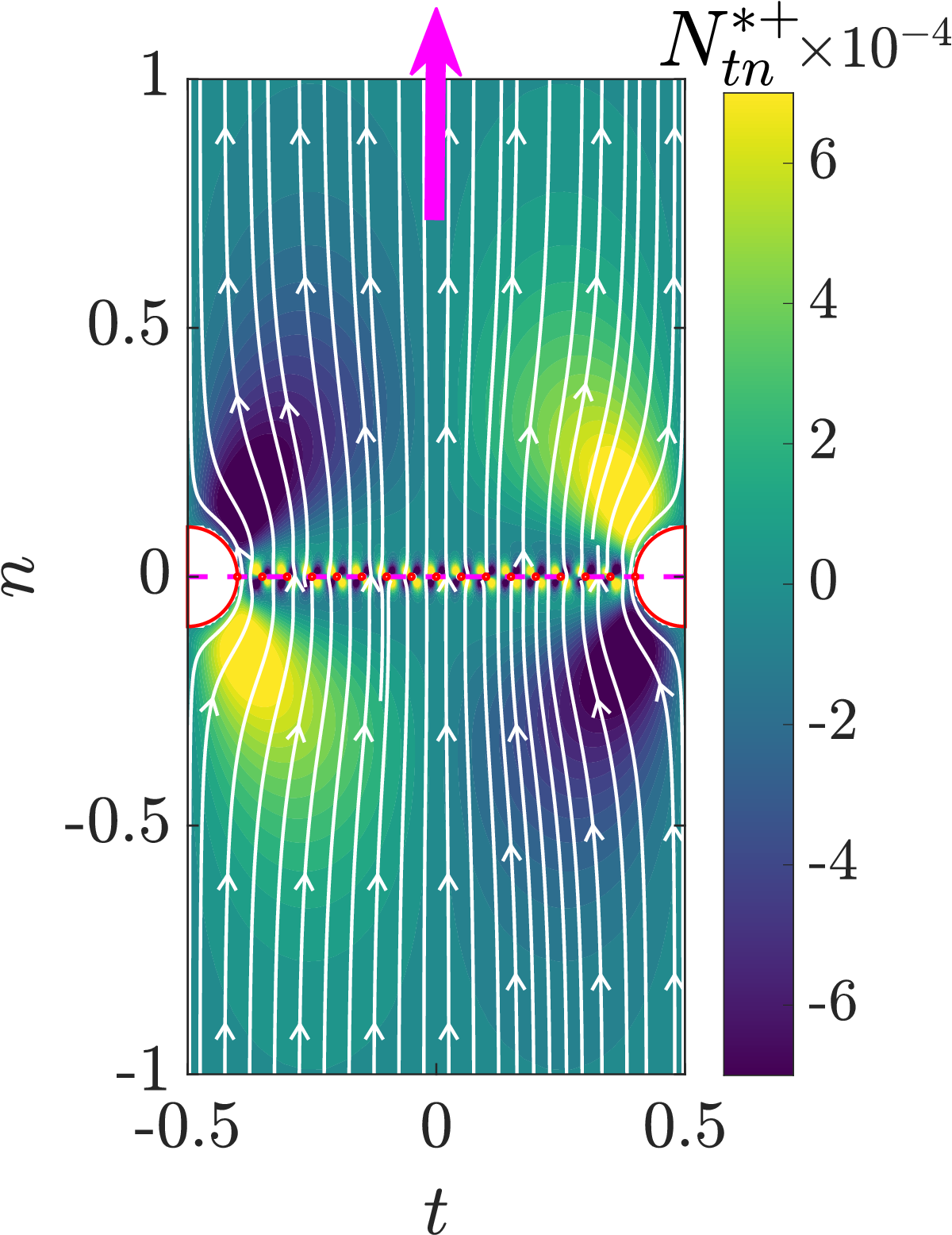}\qquad
\includegraphics[height=5cm]{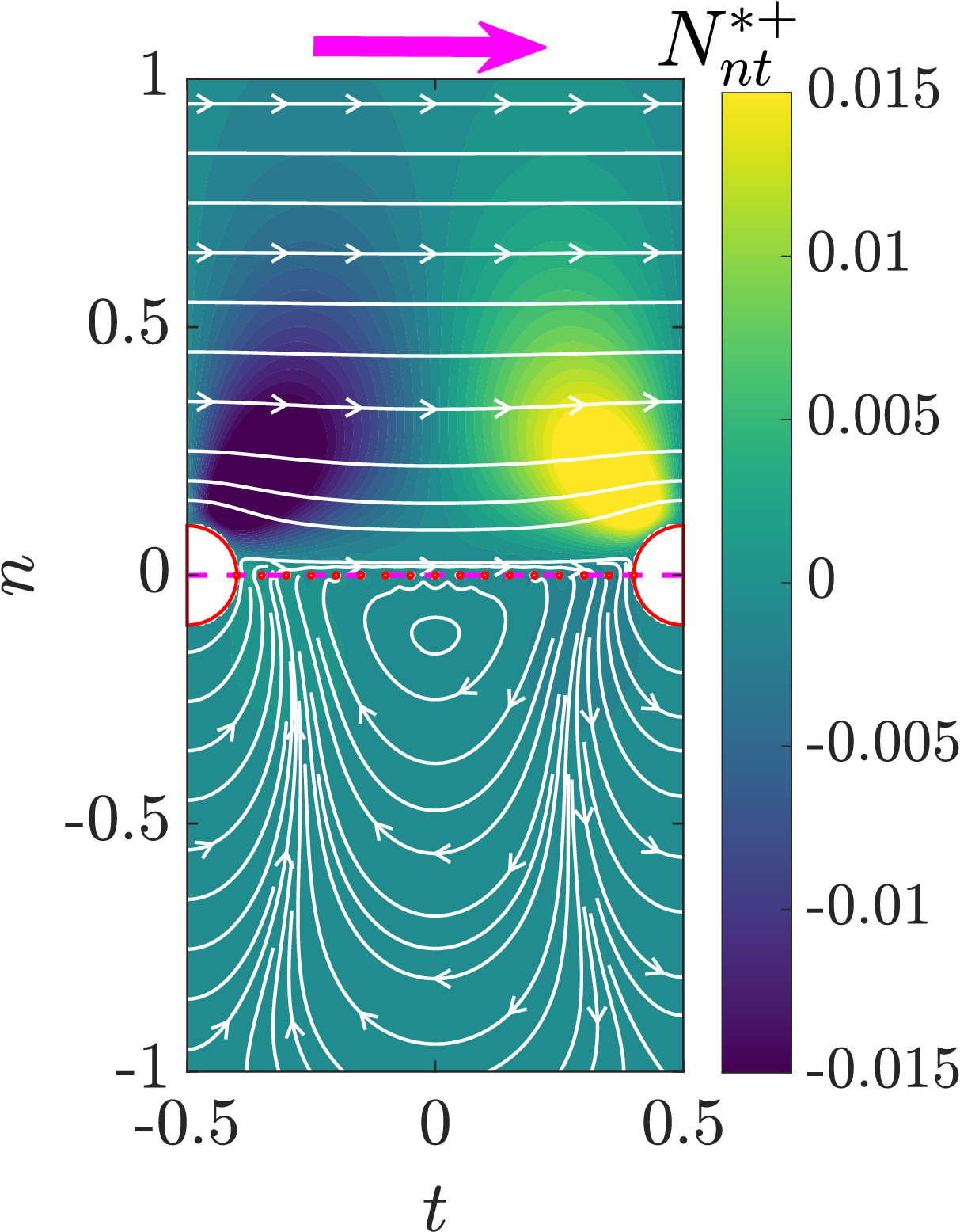}
\includegraphics[height=5cm]{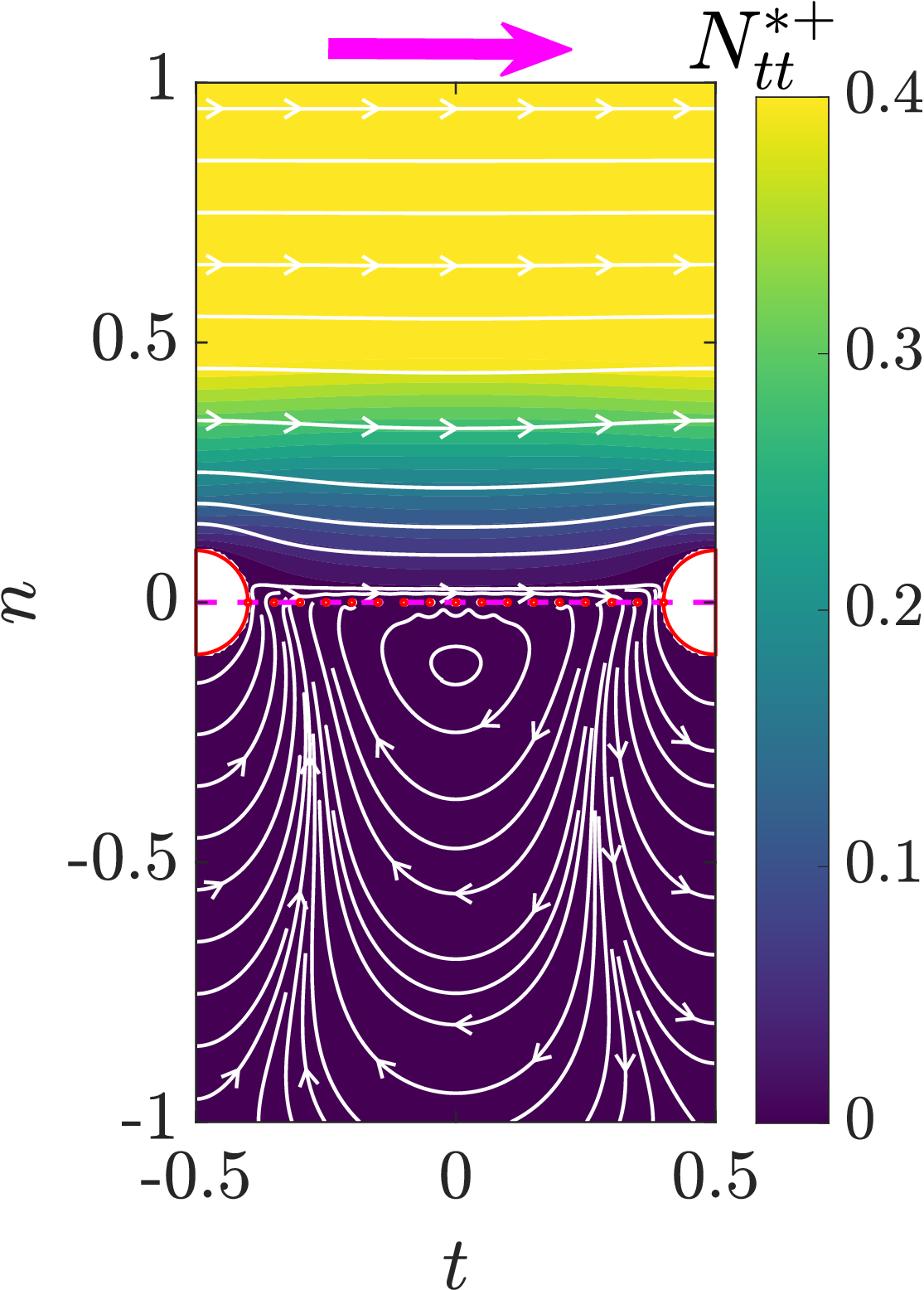}\quad
    \caption{Mesoscopic characteristic fields for the mesoscale problem {for $r_{\mathrm{micro}}=r_{\mathrm{meso}}=0.1,\delta=0.05$}.
Columns show the characteristic velocity field components (a,e) $N_{nn}^{+}$, (b,f) $N_{tn}^{+}$, (c,g) $N_{nt}^{+}$, and (d,h) $N_{tt}^{+}$, associated with a unit traction imposed on the upper side of the mesoscopic cell.
The top row corresponds to the reduced mesoscopic problem in which the inner porous membrane is replaced by the pore-scale homogenized interface law, with values inherited from the microscopic problem, Fig.~\ref{fig:microscale_characteristic_fields}.
The bottom row corresponds to the reference calculation in which the inner membrane microstructure is explicitly resolved.
Colormaps represent the velocity component and white streamlines show the characteristic velocity field associated with the same unit macroscopic forcing (magenta arrows).}
    \label{fig:mesoscale_characteristic_fields}
\end{figure}

\subsection{Microscale and mesoscale problems}

As a first step, we solve the microscale and mesoscale characteristic problems, whose characteristic geometries are shown in Fig. \ref{fig:validation_geometries}(a).  {In the considered case, the membrane is formed by two nested arrays of solid cylinders. Being $\ell'$ and $\ell$ the spacing between micro- and mesoscopic cylinders, the radii of the corresponding cylindrical inclusions are $r_{\mathrm{micro}}=0.1$ and $r_{\mathrm{meso}}=0.1$, with $\delta=0.05$, i.e. $\ell'=0.05\ell$. In the mesoscopic domain, lengths are normalized by $\ell$, so that $\varepsilon$ is set only at the scale by specifying $L$.} 
{These Stokes-like problems are forced by a boundary stress on the top or bottom far-field boundary, with periodic boundary conditions on the sides. Therefore, they are implemented using the same finite-element framework described in Section~II. {We evaluate the corresponding far-field membrane responses; formally, their domains extend to infinity in the normal direction $x_n$. However, their solution reaches asymptotically converged averages for large values of $|\mathcal{B}^--\mathcal{B}^+|$ and $|\mathcal{B}^{MESO^-}-\mathcal{B}^{MESO^+}|$. In our computational framework, we verified that $|\mathcal{B}^--\mathcal{B}^+|=4\ell'$ and $|\mathcal{B}^{MESO^-}-\mathcal{B}^{MESO^+}|=4\ell$ is sufficient to obtain converged microscopic and mesoscopic averages with an accuracy of the order of $0.1\%$.} A mesh spacing of $\Delta l_1 = 0.0125$ at the boundaries of the microscopic cell, with at least 25 grid points on the solid inclusions, is employed. Other simulations have been carried out on coarser meshes with spacing $\Delta l_2 = 0.025$ and $\Delta l_3 = 0.05$, and numerical convergence of the average values of the non-zero permeability tensor entries up to 0.1 \% has been verified between $\Delta l_1$ and $\Delta l_2$.}

The characteristic microscopic flow through the nested membrane element (Fig.~\ref{fig:microscale_characteristic_fields}) is analogous to the configuration discussed in \cite{ZampognaGallaire2020JFMStressJump}. Panels~(a,b) show the characteristic velocity field $M_{jn}^+$ ($j=t,n$), where the first (resp.\ second) component denotes the $t$ (resp.\ $n$) velocity component, for a unit normal stress across the membrane. The flow accelerates through the narrow gap and exhibits nonzero $t$ components due to the successive contraction and expansion of the flow. Panels~(c,d) instead show the components $M_{jt}^+$ ($j=t,n$) of the characteristic velocity field induced by a tangential stress applied on one side. In this case, the flow exhibits a recirculation in the vicinity of the constriction. The tangential velocity also decreases sharply to zero, with extremely small values on the opposite side of the membrane.
These fields are averaged (see Section II) to obtain the effective microscopic permeability tensor entries. By symmetry, the only nonzero components are
$\langle M_{nn}\rangle$ and
$\langle M_{tt}\rangle$.

\begin{table}[]
    \centering
    \begin{tabular}{c|c|c|c|c|c} 
        $\langle {M}_{nn} \rangle$  & $\langle {M}_{tt}\rangle$ & $\lbrace N_{nn}\rbrace$ & $\lbrace N_{tt} \rbrace$ & $\lbrace N_{nn}^{*} \rbrace$ & $\lbrace N_{tt}^{*}\rbrace $\\
$7.94 \times 10^{-2}$ & $2.76 \times 10^{-4}$
& $2.94 \times 10^{-3}$ & $5.29 \times 10^{-4}$
& $2.98 \times 10^{-3}$ & $5.36 \times 10^{-4}$ \\
    \end{tabular}
    \caption{Numerical values of the non-zero averages of the microscopic and mesoscopic tensor components ${M}_{ij}^\pm$ and $N_{ij}^\pm$ for the symmetric membrane case presented in Figs. {3 and 4}. Quantities $\lbrace N_{ij}^{*}\rbrace$ represent the components of the mesoscopic tensor $\lbrace N_{ij}\rbrace $ computed using a fully resolved mesoscopic model, where the internal solid inclusions are explicitly discretized in the mesh (see the second row of panels in Fig.~\ref{fig:mesoscale_characteristic_fields}).}
    \label{tab:tab1}
\end{table}

At the mesoscopic scale, these permeability values are prescribed on the membrane embedded in the mesoscopic domain, see Fig.~\ref{fig:mesoscale_characteristic_fields} for the characteristic velocity fields ${{N}}_{ij}^+$. The nested membrane causes a marked reduction of the velocity in its immediate vicinity, thereby affecting the pressure drop across the pits. The characteristic fields obtained with the homogenized membrane condition compare well with those where the membrane is explicitly resolved. 
The resulting microscopic and mesoscopic permeability-tensor components are reported in Table~\ref{tab:tab1}. {
For the reference configuration, the relative differences between the
nested-homogenized and fully resolved mesoscale coefficients are about
$1.3\%$ for both $\lbrace N_{nn}\rbrace$ and
$\lbrace N_{tt}\rbrace$.} The agreement between the homogenized membrane description and the resolved simulations is consistent with previous studies~\cite{ZampognaGallaire2020JFMStressJump,Ledda_Boujo_Camarri_Gallaire_Zampogna_2021}.

\begin{figure*}[t]
\centering

\begin{subfigure}{0.32\textwidth}
    \centering
    \includegraphics[width=\linewidth]{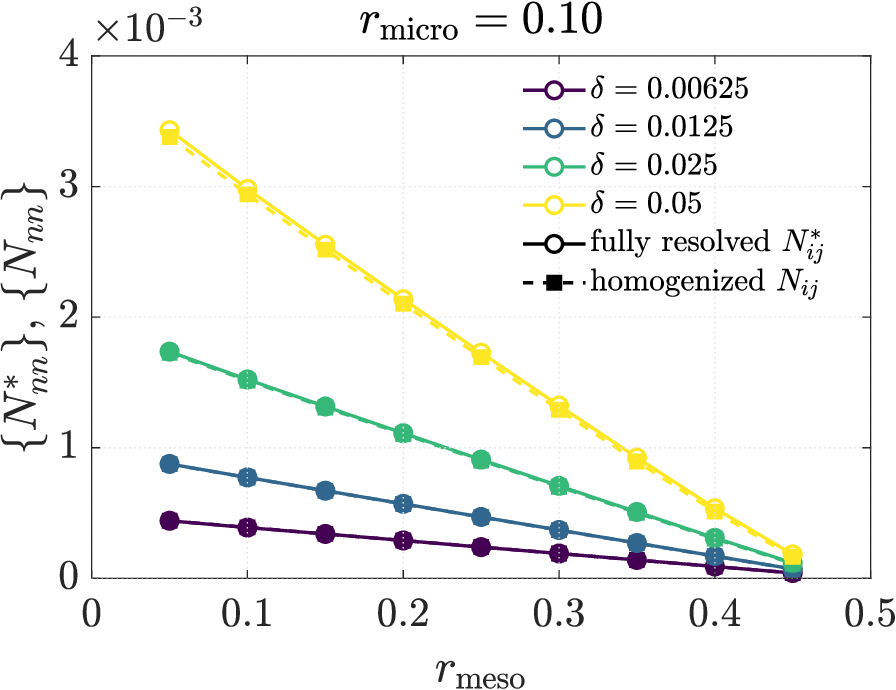}
    \label{fig:Nnn_rmicro_010}
\end{subfigure}
\hfill
\begin{subfigure}{0.32\textwidth}
    \centering
    \includegraphics[width=\linewidth]{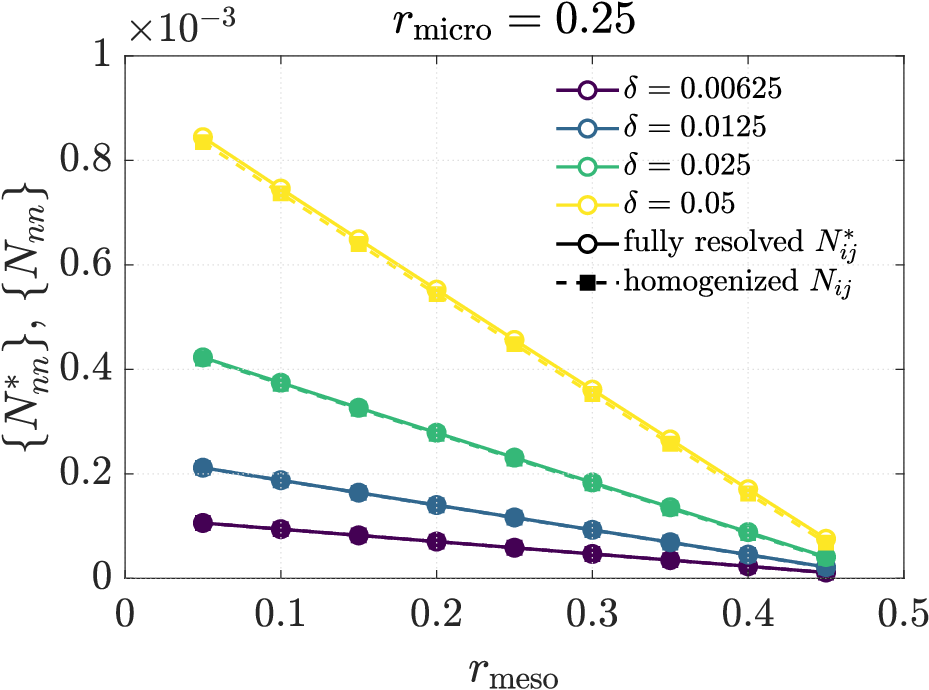}
    \label{fig:Nnn_rmicro_025}
\end{subfigure}
\hfill
\begin{subfigure}{0.32\textwidth}
    \centering
    \includegraphics[width=\linewidth]{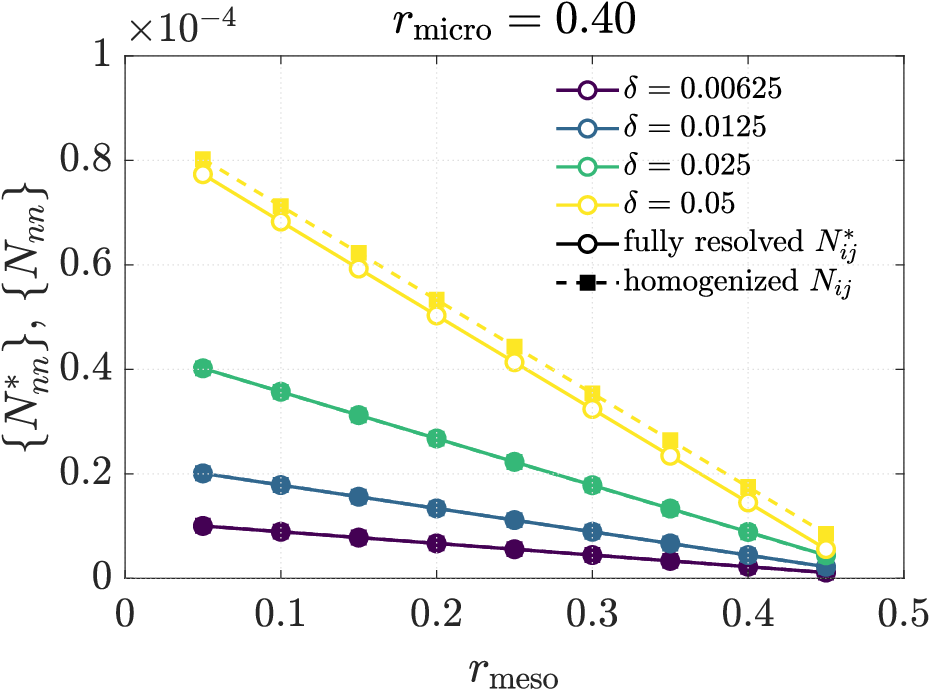}
    \label{fig:Nnn_rmicro_040}
\end{subfigure}

\vspace{0.5em}

\begin{subfigure}{0.32\textwidth}
    \centering
    \includegraphics[width=\linewidth]{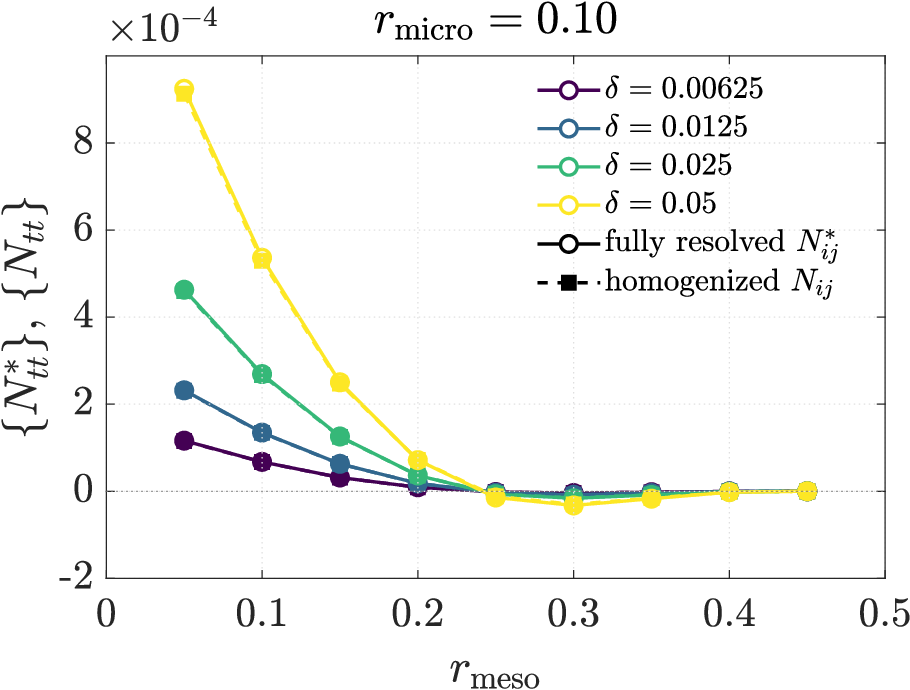}
    \label{fig:Ntt_rmicro_010}
\end{subfigure}
\hfill
\begin{subfigure}{0.32\textwidth}
    \centering
    \includegraphics[width=\linewidth]{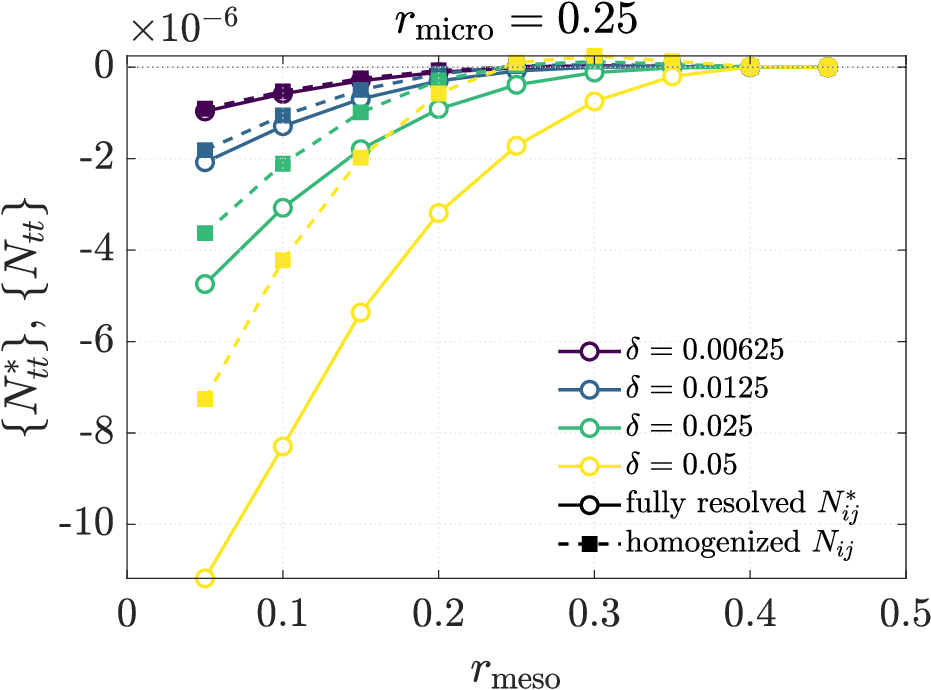}
    \label{fig:Ntt_rmicro_025}
\end{subfigure}
\hfill
\begin{subfigure}{0.32\textwidth}
    \centering
    \includegraphics[width=\linewidth]{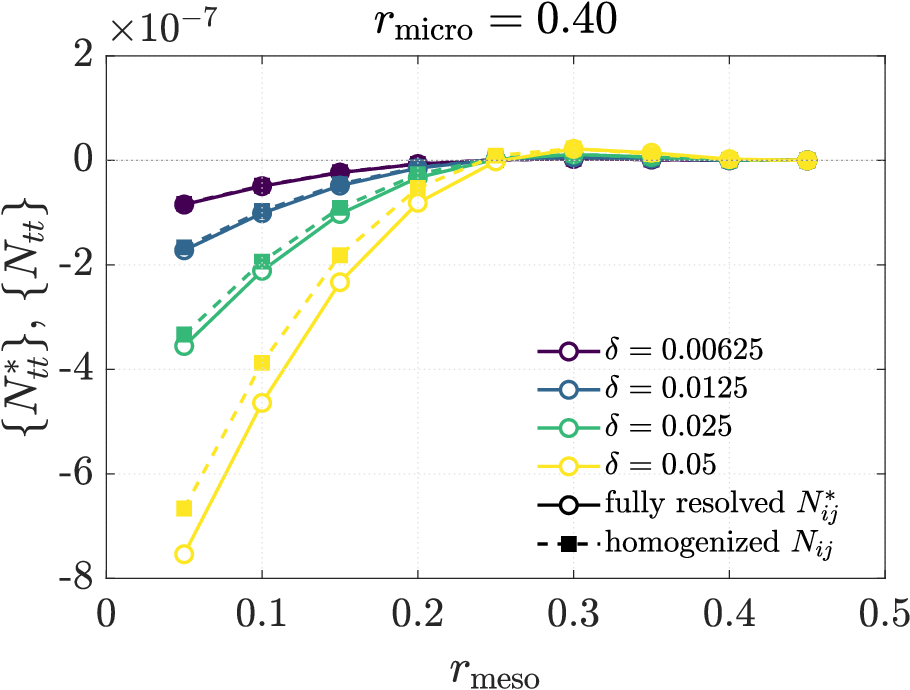}
    \label{fig:Ntt_rmicro_040}
\end{subfigure}
\caption{
{Comparison between the fully resolved coefficients
\(\lbrace N_{ij}^{*}\rbrace\) and the nested-homogenized coefficients
\(\lbrace N_{ij}\rbrace\) as functions of \(r_{\mathrm{meso}}\).
The columns correspond to different values of \(r_{\mathrm{micro}}\);
the top and bottom rows show \(\lbrace N_{nn}\rbrace\) and \(\lbrace N_{tt}\rbrace\),
respectively. Different colors denote different values of \(\delta\).
}
}
\label{fig:tensor_coefficients}
\end{figure*}


{
Fig.~\ref{fig:tensor_coefficients} compares the fully resolved mesoscale
coefficients, \(\lbrace N_{ij}^{*}\rbrace\) with those obtained using the
nested-homogenized membrane model, \(\lbrace N_{ij}\rbrace\), for different
values of the microscopic radius  \(r_{\mathrm{micro}}\) (nondimensionalized with the microscopic length), mesoscopic radius \(r_{\mathrm{meso}}\)  (nondimensionalized with the mesoscopic length) of the inclusions, and \(\delta\).
In all cases, the nonzero nested homogenization tensor entries follow the fully
resolved results.
For fixed \(r_{\mathrm{micro}}\), the magnitude of
\(\lbrace N_{nn}\rbrace\) decreases approximately linearly with
\(r_{\mathrm{meso}}\). This behavior is consistent with the progressive
reduction of the open mesoscopic cross-section available for normal flow.
Furthermore, the curves obtained for different values of \(\delta\) have nearly
the same shape and are approximately proportional to \(\delta\), indicating
that it mainly rescales the tensor components. The microscopic
geometry (or, equivalently, the microscopic porosity) has a stronger nonlinear effect on the normal permeability. 
}
{
The tangential component \( \lbrace N_{tt} \rbrace \) exhibits a non-monotonic behavior with
\(r_{\mathrm{meso}}\). First, at low \(r_{\mathrm{meso}}\), it can present positive or negative values for the same microscopic radius. As the mesoscopic radius increases, \( \lbrace N_{tt} \rbrace \) approaches zero, changes sign, reaches a small extreme of opposite sign, and finally
returns toward zero. 
An increase in \(r_{\mathrm{micro}}\) strongly reduces the maximum magnitude of
\(\lbrace N_{tt}\rbrace\) by approximately three orders of
magnitude between \(r_{\mathrm{micro}}=0.10\) and \(0.40\). Therefore,
\(r_{\mathrm{meso}}\) mainly controls the effective (tangential) slip. 
These results identify complementary roles for the two geometrical levels.
The microscopic geometry sets the intrinsic permeability of the inner membrane
and the micro-meso separation of scales parameter $\delta$ tunes linearly the overall scale of the two nonzero tensor entries. The
mesoscopic geometry instead can be used to set the effective slip values, while producing a comparatively weaker variation of the normal component.  The nested architecture
therefore provides a largely independent geometrical control of the normal and tangential effective responses by combining the two geometrical degrees of freedom and the two separation of scale parameters.
}

\subsection{Macroscale problem - tracheid-inspired connection}
\begin{figure}[t!]
\centering
\includegraphics[width=\linewidth]{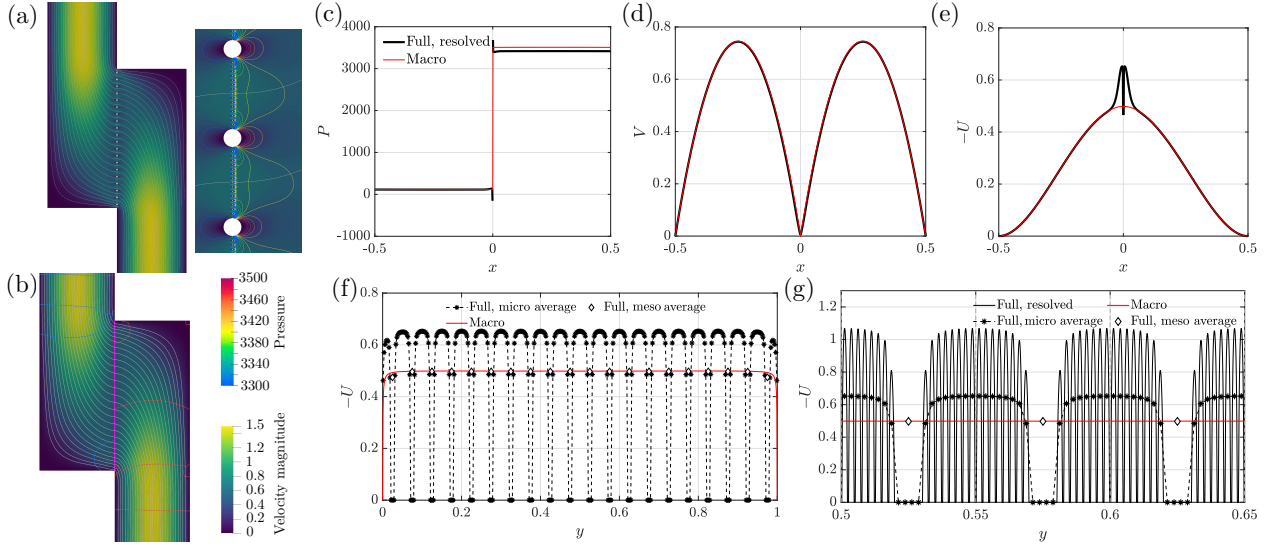}
\caption{{Tracheid-like connection with $r_{\mathrm{micro}}=r_{\mathrm{meso}}=0.1,\delta=0.05,\varepsilon=0.05$}. Comparison between the fully resolved and homogenized descriptions in a two-dimensional tracheid-like connection. (a,b) Colormaps of the velocity magnitude overlaid with streamlines (white lines) and selected iso-levels of the pressure field: (a) fully resolved solution, where both mesoscopic pits and microscopic pores are explicitly represented, together with a zoom close to the membrane; (b) homogenized solution, where the nested porous region is replaced by an effective zero-thickness interface law. (c--e) Profiles extracted along the line $y=0.5$, orthogonal to the membrane: (c) pressure, (d) velocity component parallel to the membrane, and (e) velocity component normal to the membrane, for fully resolved (black lines) and homogenized (red lines) simulations. {(f,g) Normal-to-the-membrane velocity sampled along the membrane. Panel (f) compares the microscopic (stars) and mesoscopic (diamonds) averages of the full-scale solution with the homogenized prediction (red line). Panel (g) additionally shows the fully resolved profile (black line) in a zoomed region, together with the two averaged full-scale quantities and the homogenized prediction.}}
\label{fig:tracheid_connection_validation}
\end{figure}

We next proceed to the tracheid-like macroscopic problem, Fig.~\ref{fig:validation_geometries}(b). In the macroscale nondimensionalization, the non-zero entries of the permeability tensor read $\mathcal{N}_{nn}=\varepsilon \lbrace N_{nn} \rbrace = 1.47 \times 10^{-4}$ and $\mathcal{N}_{tt}= 2.65\times 10^{-5}$.
{Regarding the numerical implementation, a spatial convergence study was performed both for the full-scale case and for the macroscopic solution. For the fully resolved case, convergence within $1\%$ in the average pressure difference across the membrane (evaluated between two lines at a distance of $0.05L$ on each side) is obtained by successively halving the element size at the solid inclusions, with at least $20$ points along each solid inclusion. This leads to meshes containing approximately {\(2\)--\(10.5\times 10^{6}\) elements, depending on the considered case (see Table \ref{tab:pressure_drop_comparison} for the considered values of $\varepsilon$ and $\delta$).} Since micro- and meso-scale geometrical features are not resolved in the macroscopic model, far less restrictive mesh requirements are sufficient. Numerical convergence of the fluid velocity components at the membrane is achieved with a mesh spacing of $\Delta l_1 = 0.025L$ (about 20$\,$000 elements), yielding differences in the pressure drop across the membrane of less than 0.1$\%$.}
Fig.~\ref{fig:tracheid_connection_validation}(a) reports the velocity and pressure from the full-scale simulation. As expected, the velocity magnitude decreases and the streamlines progressively bend near the porous membrane. 
The macroscopic model (panel~b) reproduces well the flow, while smoothing the localized high-pressure and velocity peaks close to individual inclusions.
Panels~(c--e) provide local comparisons of averaged pressure and velocity profiles as functions of $x$, evaluated along the line $y=0.5$ that passes through the membrane mid-point. Sampled quantities agree with the full-scale results, with discrepancies on the order of the microscopic and mesoscopic unit-cell sizes. 
Finally, {panels~(f,g)} show the normal velocity sampled along the membrane surface. {As highlighted by the zoom in panel~(g),} in the full-scale simulation, the flow within each microscopic pore exhibits an approximately parabolic profile. The envelope of these pore-scale profiles forms, at the mesoscopic level, a Poiseuille-like distribution. After mesoscopic averaging, the full-scale results agree with the macroscopic predictions.

\begin{table}[t]
\centering
\begin{tabular}{lccc}
\hline
Case
& $\Delta P_\mathrm{full}$
& $\Delta P_\mathrm{hom}$
& Relative error \\
\hline
$\varepsilon_0,\delta_0$
    & $3.30\times10^{3}$
    & $3.39\times10^{3}$
    & $2.7\%$ \\

$\varepsilon_0/2,\delta_0$
    & $6.56\times10^{3}$
    & $6.78\times10^{3}$
    & $3.4\%$ \\

$2\varepsilon_0,\delta_0$
    & $1.66\times10^{3}$
    & $1.70\times10^{3}$
    & $2.4\%$ \\

$\varepsilon_0,\delta_0/2$
    & $6.45\times10^{3}$
    & $6.57\times10^{3}$
    & $1.9\%$ \\

$\varepsilon_0/2,\delta_0/2$
    & $1.28\times10^{4}$
    & $1.31\times10^{4}$
    & $2.3\%$ \\

$2\varepsilon_0,\delta_0/2$
    & $3.23\times10^{3}$
    & $3.29\times10^{3}$
    & $1.9\%$ \\
\hline
\end{tabular}
\caption{Tracheid-like connection: comparison of the mean pressure
drop $\Delta P$, measured along two lines parallel to the membrane and of the same length, at a normal distance of $\pm 0.05$, between the full-scale simulations and the homogenized
model for different scale-separation parameters obtained from the
baseline $\varepsilon_0=0.05$ and $\delta_0=0.05$ of
Fig.~\ref{fig:tracheid_connection_validation}. The relative error is
computed with respect to the full-scale result as
$100|\Delta P_\mathrm{hom}-\Delta P_\mathrm{full}|/
\Delta P_\mathrm{full}$.}
\label{tab:pressure_drop_comparison}
\end{table}

{
We conclude this section by exploring the effect of the two scale-separation parameters, $\varepsilon_0$ and $\delta_0$, on the pressure drop, measured along two lines at a distance of $0.05L$ from the membrane. As shown in Table~\ref{tab:pressure_drop_comparison}, the homogenized model reproduces both the magnitude and the parametric trends of the full-scale simulations. In particular, at fixed $\delta_0$, $\Delta P$ approximately doubles when $\varepsilon_0$ is halved, and approximately halved when $\varepsilon_0$ is doubled. A similar increase is observed when $\delta_0$ is halved. The relative error remains of the order of the larger separation of scale parameters, with the homogenized model consistently producing a slightly larger pressure drop than the full-scale simulations. Overall, the homogenized formulation is robust with respect to changes in the relative dimensions of the nested structure and preserves the effective hydraulic response of the resolved geometry nearly across one order of magnitude, for the present case.
}

In summary, this first test case shows how the proposed nested multiscale framework can capture the dominant macroscopic flow features in a two-dimensional connected geometry. The macroscopic model reproduces the global pressure drop and the redistribution of velocity across the membrane. While local microscopic and mesoscopic variations are necessarily filtered out by the averaging procedure, their cumulative effect is taken into account through the effective permeability properties.
In addition, the computational requirements are substantially lower. The fully resolved benchmark requires a number of elements of the order $10^{7}$ to resolve simultaneously the mesoscopic inclusions and the microscopic membrane pores, whereas the homogenized model uses approximately $2\times 10^{4}$ elements for the corresponding macroscopic computation. Thus, the proposed interface closure reduces the mesh complexity by almost three orders of magnitude while retaining the leading-order pressure drop and membrane-averaged flow redistribution. This computational gain is essential for parametric studies and for the network-scale simulations.

\begin{figure}[t!]
    \centering
    \includegraphics[width=0.9\linewidth]{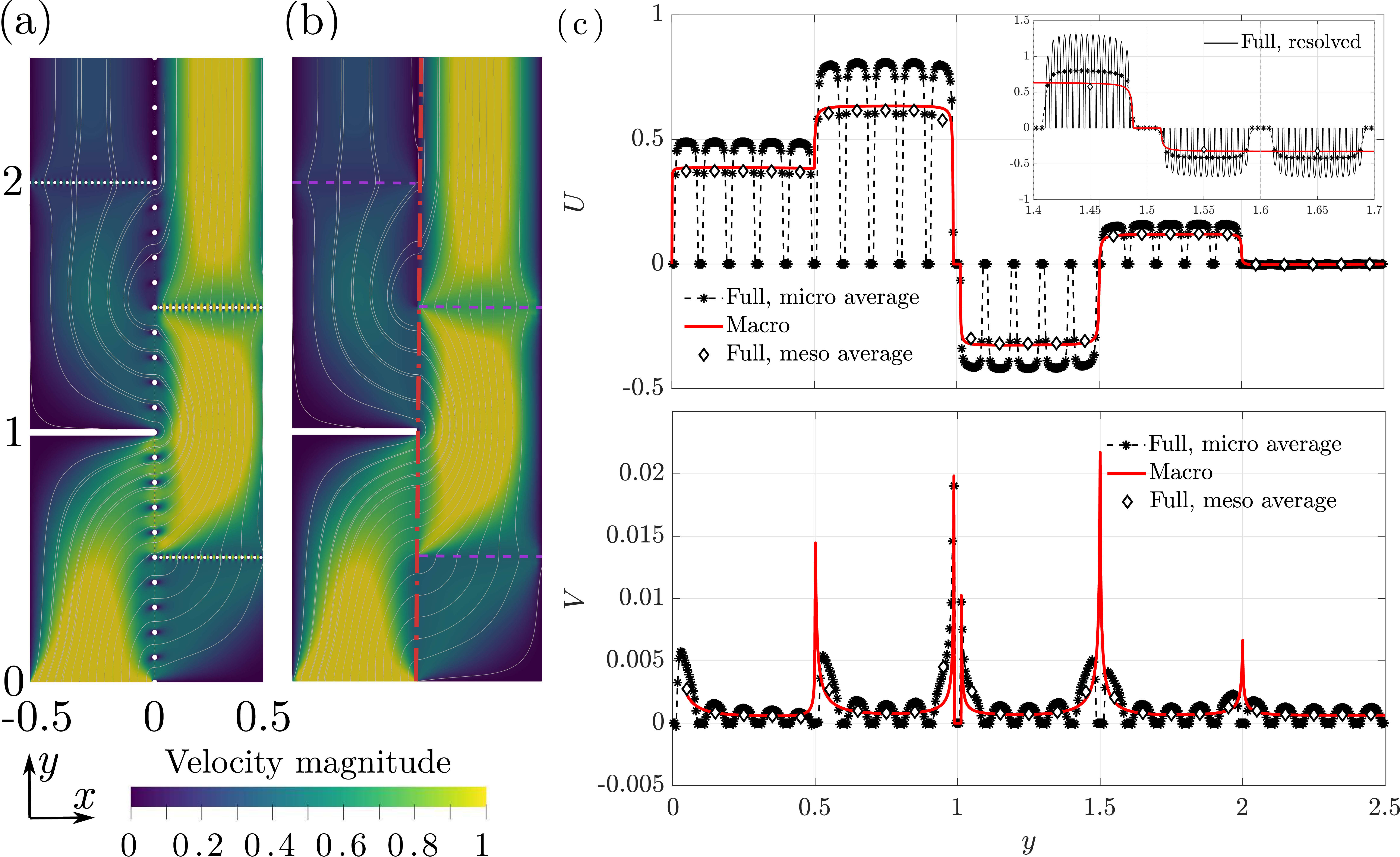}
    \caption{Comparison between the fully resolved and homogenized descriptions in a two-dimensional vessel-like connection for nested membranes with $r_{\mathrm{micro}}=r_{\mathrm{meso}}=0.1,\delta=0.05,\varepsilon=0.1$ {and transverse obstructions with radius $r_{transv}=0.25$ and scale separation parameter $\varepsilon_{transv}=0.025$}.
(a) Fully resolved solution, where the mesoscopic pit array and microscopic membrane pores are explicitly represented.
(b) Homogenized solution, where the nested porous region is replaced by an effective zero-thickness interface law.
In panels (a,b), colormaps show the velocity magnitude and white lines denote streamlines.
(c) Velocity components normal ($U$, top) and tangential ($V$, bottom) to the membrane, sampled along the nested permeable interface. The black solid line denotes the fully resolved solution; stars (connected by dashed lines) and diamonds denote microscopic and mesoscopic averages of the full-scale solution, respectively, and the red line denotes the homogenized macroscopic prediction. The inset in panel (c) provides a zoom of the normal-velocity profile.}
\label{fig:vessel_connection_global_validation}
\end{figure}

\subsection{Macroscale problem - vessel-like connection}

A second, complementary, two-dimensional configuration, inspired by lateral exchange between adjacent vessels, is briefly presented. In contrast with the tracheid-like geometry discussed above, where the upstream conduit terminates at the membrane edge, here two neighboring conducting channels run side by side and communicate through a finite permeable region along their common wall (Fig.~\ref{fig:validation_geometries}c). An impervious barrier is located just downstream of the inlet of the left channel. This configuration provides a distinct test of the homogenized interface law since the flow can redistribute between the two channels depending on the local hydraulic resistance of the nested porous interface. 
From a numerical point of view, the full-scale model is resolved using the same mesh-convergence criterion as in the previous case, leading to a mesh composed of about $5\times10^7$ elements. The non-zero tensor entries for the macroscopic model are $\mathcal{N}_{nn}=2.94\times10^{-4}$ and $\mathcal{N}_{tt}=5.29\times10^{-5}$ for the vertical nested membrane ($\varepsilon=0.1$, $\delta=0.05$), and {$\mathcal{N}_{nn}^{p}=4.7\times10^{-4}$ and $\mathcal{N}_{tt}^{p}=1.3\times10^{-3}$}, obtained from a single-scale homogenization problem~\cite{ZampognaGallaire2020JFMStressJump} with a geometry that has the same topology as the one in Fig.~\ref{fig:microscale_characteristic_fields}, but with a different nondimensional radius $r_{transv}=0.25$.

Fig.~\ref{fig:vessel_connection_global_validation} compares the fully resolved and homogenized solutions. The flow deviates from the left channel to the right channel, redistributing upstream and downstream the first perforation plate of the right channel. Subsequently, the flow follows mainly the right channel, although some leakage toward the left channel is observed. This residual leakage provides a test of the interface model, since the homogenized description must capture not only the relation between main bypass flow and pressure drop but also weaker cross-channel exchange. As shown in Fig.~\ref{fig:vessel_connection_global_validation}, the macroscopic model reproduces the global flow organization and the exchange region between the two channels, while avoiding the explicit resolution of each nested pore.
Panel~(c) provides a local comparison along the nested permeable interface. 
The mesoscopic averages closely follow the
macroscopic prediction along the different membranes.
The tangential component is substantially smaller in magnitude and presents localized variations associated with the flow redistribution between the two vessels. Its averaged behavior is also captured by the homogenized description. Therefore, the effective interface law also reproduces the weaker tangential response along the membrane. Together with the tracheid-like benchmark, this second test confirms that the nested closure remains accurate across distinct two-dimensional xylem-inspired connectivity patterns.
Furthermore, the pressure redistribution between the two vessels provides an additional
quantitative assessment of the reduced model. As reported in
Table~\ref{tab:vessel_pressure_difference}, the homogenized solution
reproduces the pressure difference (left side minus right side) between the two channels in three
parts of the membrane. {According to the inlet position and under the influence of the impermeable left transverse obstruction, the fluid is forced to flow from left to right, resulting in a positive pressure difference. Immediately downstream of the impermeable obstacle, in the right part of the vessel, the fluid encounters a perforated plate whose associated resistance partially redirects the fluid to the left part of the vessel, resulting in a negative pressure difference for $1<y<1.5$. The same mechanisms govern the flow behavior for $1.5<y<2$, where the pressure difference becomes positive again.} The relative difference between homogenized model and fully resolved simulations is of the order of a few percent.

\FloatBarrier

\section{Network-scale response of a porous vascular network}

\begin{table}[!t]
\centering
\resizebox{0.8\textwidth}{!}{%
\begin{tabular}{ccc|ccc|ccc}
\hline
 \multicolumn{3}{c}{$\Delta P_{0.5 \rightarrow 1}$}
& \multicolumn{3}{c|}{$\Delta P_{1 \rightarrow 1.5}$}
& \multicolumn{3}{c|}{$\Delta P_{1.5 \rightarrow 2}$}\\
 Full & Hom. & Diff.
& Full & Hom. & Diff.
& Full & Hom. & Diff. \\
\hline
 $2.04\times10^{3}$ & $2.10\times10^{3}$ & $2.9\%$
& $-1.05\times10^{3}$ & $-1.07\times10^{3}$ & $1.9\%$
& $3.92\times10^{2}$ & $3.88\times10^{2}$ & $1.0\%$ \\
\hline
\end{tabular}%
}
\caption{Vessel-like connection: comparison between the full-scale and
homogenized average pressure differences (left side minus right side) between two lines located at a normal distance of {0.01} from the membrane for each part {upstream of the impervious barrier ($0.5<y<1$), at the mid-height ($1<y<1.5$), and right upstream of the last transverse porous barrier ($1.5<y<2$).} The relative differences are computed with respect
to the corresponding full-scale results.}
\label{tab:vessel_pressure_difference}
\end{table}

\begin{figure}
    \centering
    \includegraphics[width=\linewidth]{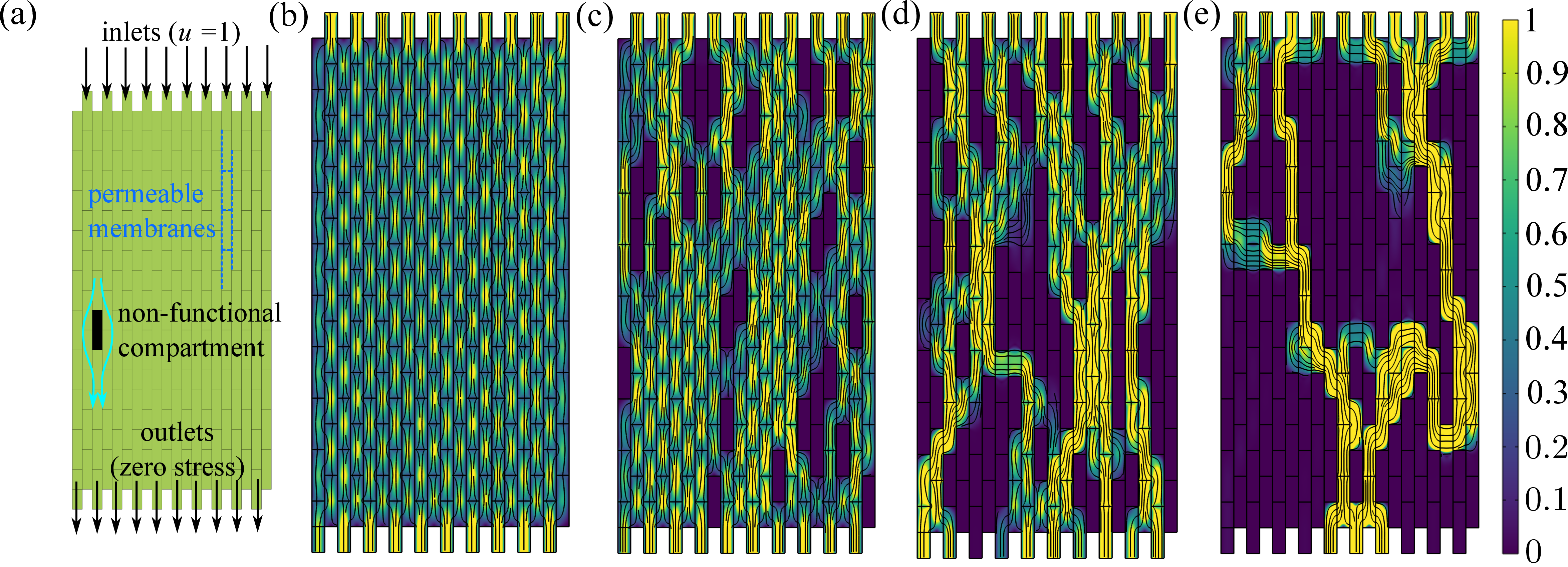}
    \caption{Xylem-like network geometry and representative velocity fields under progressive loss of functionality. (a) Schematic of the staggered array. (b-e) Velocity magnitude colormaps and streamlines (black lines) (b) in the intact network, and in representative damaged configurations with damaged fractions (c) $f=16.7\%$, (d) $33.3\%$, and (e) $50.0\%$, respectively. The colormap is saturated for visualization purposes.}
    \label{fig:network_velocity_fields}
\end{figure}

We conclude our analysis by quantifying the hydraulic response of a two-dimensional xylem-inspired network composed of simplified conducting elements. The network is represented as an ideal array of staggered rectangular compartments (or cells), an extension of the vessel-like connection shown in the previous Section. Each side of the rectangle is modeled as a homogenized permeable membrane with assigned (constant) normal permeability and tangential slip properties. Adjacent compartments exchange fluid through these membrane interfaces, which are imposed through the effective stress-jump condition. 
Each rectangular xylem compartment, of dimensions $L \times 4 L$, is replicated 20 times along the transverse ($x$) direction and 10 times along the streamwise ($y$) direction. The Stokes equations are nondimensionalized with the (constant) normal velocity at each inlet $U$ and with $L$. 
We
consider the case
${\mathcal{N}}_{nn}=4.5\times10^{-5}$ and
$\mathcal{N}_{tt}=1.2\times10^{-5}$, representative of the parameter range explored above. 
{As reported in Fig.~\ref{fig:network_velocity_fields}(a), the upper boundaries are selected as inlets with prescribed unit velocity, whereas the lower boundaries are treated as zero-stress outlets. The lateral boundaries of the whole array are treated as solid walls.}

The loss of functionality of a prescribed set of cells ($\Omega_{\mathrm{NF}}$) is modelled via a Brinkman-type penalization,
\begin{equation}
-\nabla p+\nabla^2\bm{U}-\alpha(\bm{x})\bm{U}=0,
\qquad
\nabla\cdot\bm{U}=0, \quad
\alpha(\bm{x})=
\begin{cases}
\alpha_{\mathrm{off}}= 10^{8}\gg 1, & \bm{x}\in\Omega_{\mathrm{NF}},\\
0, & \mathrm{otherwise}.
\end{cases}
\label{eq:penalized_stokes}
\end{equation}
The penalization drives $\bm{U}\to \mathbf{0}$ inside these non-functional cells and therefore mimics their loss of hydraulic functionality. The Brinkman penalization in non-functional cells is imposed in weak form as a source contribution.

Simulations are performed over random sets of non-functional cells.
One realization consists of selecting $n$ random distinct cells to form this set ($\Omega_{\mathrm{NF}}$), solving the penalized Stokes problem, and extracting the corresponding hydraulic response. Repeating this procedure $n_r=70$ times yields a distribution of hydraulic responses associated with that damage level.
As before, the numerical implementation relies on a mesh composed of third-order Lagrange velocity and discontinuous first-order Lagrange pressure quadrilateral elements, whose size close to the membrane $0.05L$ is chosen after a mesh convergence analysis on a subset of 10 realizations for all defects (approximately 300$\,$000 elements). Specifically, when decreasing the element size down to $0.035L$ and $0.02L$, the pressure drop difference is monotonic with the refinement and increases of at most $0.3\%$.
Since the network is driven at fixed imposed inlet flow rate, the output of each realization is the pressure drop $\Delta P=P_{\mathrm{inlet}}-P_{\mathrm{outlet}}$ required to sustain that imposed flow rate, obtained by averaging the pressures at each inlet and outlet. We denote by $\Delta P_0$ the pressure drop of the intact network and by $\Delta P_f^{(r)}$ the pressure drop of the $r$-th realization at damaged fraction $f=n/n_{\mathrm{tot}}$, where $n_{\mathrm{tot}}=180$ is the total number of possibly non-functional cells, excluding 10 inlet and 10 outlet ones, of the xylem-like network. 

For each set, the two-dimensional network was first interpreted as an undirected graph, built directly from the known geometry of the staggered array, where vertices represent conducting compartments while edges represent hydraulic communication through shared homogenized membrane interfaces. For a random realization, the corresponding accessible graph was defined as the vertex-induced subgraph, i.e. non-functional cells and all edges whose endpoint belongs to those vertices were removed in the topological test. The set was exploited for the realization only if every inlet vertex belonged to a connected component containing at least one outlet vertex, to ensure well-posedness of the pressure-drop calculations. Therefore, statistics are conditional on the connectivity. This procedure has been implemented in MATLAB, and feeds directly the COMSOL model by selecting only random sets that satisfy this condition and on which the Brinkman penalization is imposed. 

{Representative velocity fields are shown in Fig.~\ref{fig:network_velocity_fields}(b--e). In the intact configuration (panel b), the velocity field reflects the periodic staggering of the permeable membrane interfaces. When non-functional cells are introduced, the flow is progressively redistributed through the remaining connected pathways. At $f=16.7\%$ (panel c), inactive regions already create local low-velocity compartments and force the fluid to bypass blocked cells through neighboring membrane connections, where the mean velocity increases. At $f=33.3\%$ (panel d), the conducting paths become more tortuous and the imposed flow is carried by a smaller subset of connected regions. At $f=50.0\%$ (panel e), the response is strongly channelized: only a few inlet-to-outlet pathways carry most of the imposed flow, with localized counterflow to follow channelization. Therefore, the hydraulic response is controlled not only by the number of disabled compartments, but also by their spatial arrangement relative to the available inlet-to-outlet pathways.}

Since the problem remains linear, the hydraulic conductivity is inversely proportional to the pressure drop at fixed flow rate through the conductivity $K$:
\begin{equation}
K_{\rm rel}^{(r)}(f)
=
\frac{K^{(r)}(f)}{K(0)}
=
\frac{\Delta P_0}{\Delta P_f^{(r)}} \quad \rightarrow \quad
{PLC}^{(r)}(f)
=
100\left[1-K_{\rm rel}^{(r)}(f)\right],
\end{equation}
where ${PLC}^{(r)}(f)$ is the percentage loss of conductivity, with respect to the completely functional case, of the considered realization. 
\begin{figure}
\noindent\makebox[\linewidth]{%
  \makebox[0.5\linewidth][s]{\raggedleft (a)}%
  \makebox[0.5\linewidth][s]{\raggedleft(b)}%
}
    \includegraphics[width=0.495\linewidth]{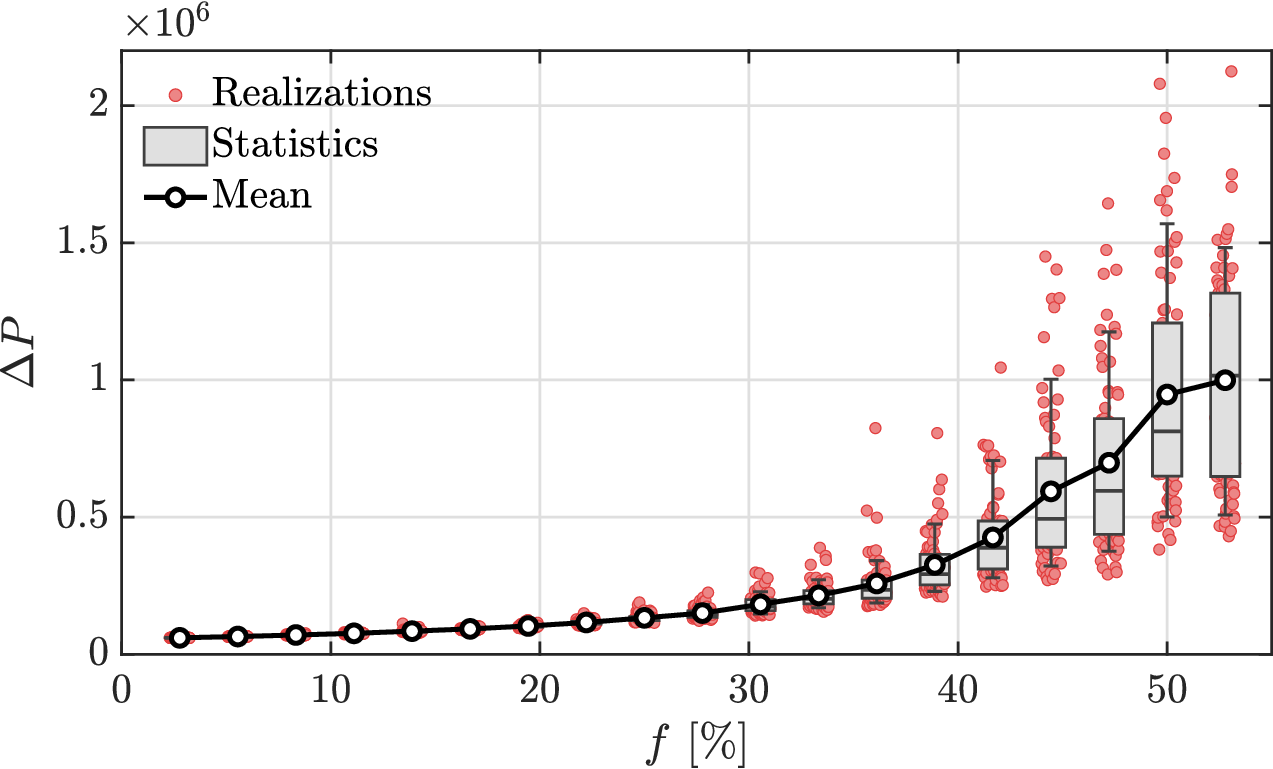}
    \includegraphics[width=0.495\linewidth]{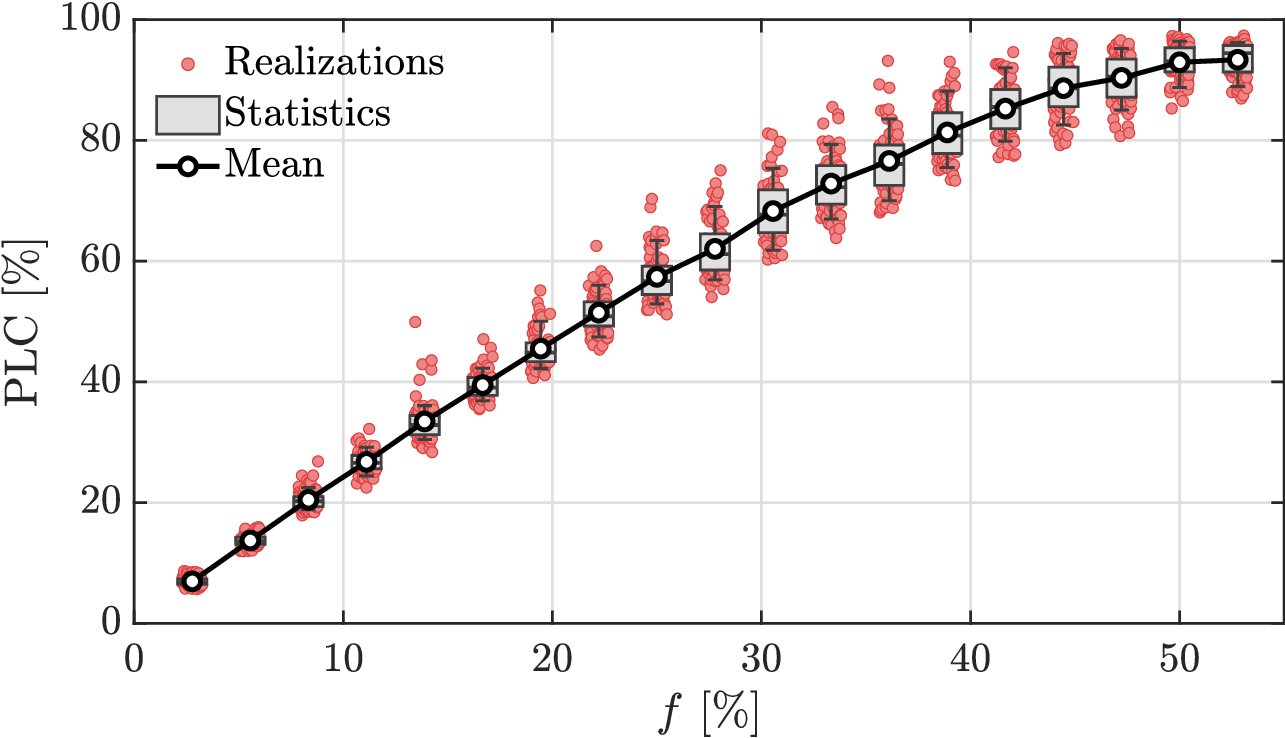}
    \caption{{Hydraulic response of the xylem-like network as a function of the damaged fraction $f$. (a) pressure drop $\Delta P$ required to sustain the imposed inlet flow rate. (b) corresponding percentage loss of conductivity, ${PLC}$. Red markers denote individual realizations, displaced horizontally by small random offsets of at most 0.5 to improve visualization. Gray boxes span the 25th–75th percentiles, the central horizontal segment indicates the median, and whiskers extend from the 10th to the 90th percentile. Black open circles connected by solid lines denote ensemble means.}}
    \label{fig:network_ensemble_response}
\end{figure}
{The mean hydraulic response is reported in Fig.~\ref{fig:network_ensemble_response}. The mean pressure drop ($\Delta P$, panel a) increases monotonically with the damaged fraction, as larger values of $\Delta P$ are required to maintain the same imposed flow rate through a progressively less connected hydraulic network. The increase spans more than one order of magnitude over the range of damaged fractions considered. The growth is initially linear with the damage, followed by an increase in the variation of $\Delta P$ with the damage and by a saturation at about $f=50\%$. The dispersion between realizations markedly grows with $f$. Therefore, once a significant number of cells is disabled, the pressure drop markedly depends on the particular spatial distribution of the non-functional cells, i.e., two configurations with the same damaged fraction may have different hydraulic costs depending on whether the remaining functional cells form efficient vertical pathways or force the flow through tortuous paths.}

{This information can be expressed in terms of percentage loss of conductivity in Fig.~\ref{fig:network_ensemble_response}(b). The central $50\%$ band and the individual realizations further highlight the dispersion of the hydraulic response. Its mean value ${PLC}$ increases approximately linearly at small and intermediate damaged fractions and then approaches a high-loss regime. The mean response reaches ${PLC}\simeq 50\%$ when approximately $20\%$ of the cells are disabled, while damage levels close to one half of the available cells lead to severe hydraulic loss, with ${PLC} > 90\%$. This amplification of local loss of functionality into macroscopic conductivity loss is in line with network-based xylem models, despite our idealization of the network topology. For example, Loepfe \emph{et al.}~\cite{Loepfe2007JTB} found that a reduction of $50\%$ in conductivity was observed with $78.1\%$ of functional conduits.}

{The homogenized network model retains the leading-order hydraulic responses observed in the simple connection of Section III and translates it into the xylem-like architecture. Although the reduced formulation necessarily filters out the local velocity peaks and pressure variations associated with individual mesoscopic and microscopic pores, it provides a computationally tractable closure for large stochastic simulations to quantify the hydraulic cost of rerouting through membrane-mediated connections. Several damaged configurations can be explored while still retaining a direct mechanistic link across scales. }

\section{Conclusion}

In this work, we developed a nested homogenization framework for porous structures characterized by multiple separated length scales. The geometrically complex nested membrane architecture is replaced by an effective interface law whose coefficients are obtained from local Stokes problems posed at the micro- and meso-scales. The hydraulic effect of the microscopic membrane pores is thus propagated systematically from the pore scale to the macroscopic scale through explicit bridges given by membrane laws and characteristic Stokes-like problems. Compared to previous developments, the effective interface obtained at the pore scale is itself embedded inside a larger characteristic problem, so that the smallest-scale response becomes the input of a second homogenization step. This produces a direct closure from pore geometry to macroscale flow redistribution. Xylem-inspired fluidic configurations were studied as a representative nested architecture to validate and illustrate the framework.

Graph-based and network-scale descriptions of xylem-like fluidic circuits have shown that connectivity and network topology strongly affect hydraulic efficiency and robustness~\cite{Loepfe2007JTB,Mrad2018PCE,Wason2021PlantPhysiol}. The present formulation provides a complementary bottom-up closure. Once the geometry of the nested porous structure is specified, the effective tensors entering the interface condition are obtained from local characteristic problems and can be transferred to the conduit- or network-scale model.
The main advantage of this strategy is twofold. It makes detailed fluid dynamics simulations of these networks computationally tractable. In the case of xylem, for example, fully resolving every connection in a large network becomes easily prohibitive, as shown in Section III where millions of elements were needed to ensure convergence in relatively small connections. The homogenized model retains the leading-order hydraulic effect of the unresolved structure through a small set of effective coefficients, at a substantially lower computational cost. We also hope that the proposed simple numerical implementation within a finite-element framework would stimulate further use of the model. Furthermore, the model preserves an explicit link between local structure and global response. The model thus provides a controlled way to test how changes at different scales propagate toward network-scale hydraulic performance.
{The idealized loss-of-function study illustrates how the use of the homogenized membrane law in a xylem-like network allows one to quantify the hydraulic cost of rerouting flow through the remaining membrane-mediated connections. The resulting hydraulic response emerges from the coupling between network topology and local hydraulic transfer.}

The present work focused on rigid nested membranes under single-phase viscous flow and employed simplified two-dimensional xylem-inspired geometries. 
Several extensions follow from this formulation. More realistic geometries and non-periodic microstructures can be incorporated by solving the same local closure problems. Furthermore, surface chemistry may affect the local transport law, especially when solving for extremely small pores. In that regime, the nested architecture could be retained while calibrating and enriching the smallest continuum characteristic problems, also with molecular-dynamics-informed closures that enter the larger-scale homogenized membrane law through effective coefficients~\cite{ChinappiZampognaMDHomogenization}.
A further extension concerns deformable channels and pit membranes. It has been extensively shown that soft elements can induce non-trivial hydraulic response in the Stokes regime~{\cite{Gomez2017,Box2020,Boyko2020,Peretz2020,Christov2022JPCM,Martinez2024,Louf2020PRL,Goncharuk2023,Oshri2024,ledda2025snapping}}.  In that case, the local geometry entering the characteristic problems is no longer fixed, but depends on the stresses imposed by the larger-scale flow. The effective tensors would therefore become dynamic quantities, and the homogenized law would have to be coupled to the membrane deformation, similar to Wittkowski \emph{et al.} \cite{WittkowskiPonteLeddaZampogna2024JFMQuasiLinear} for weak inertia effects at the pore scale. Such a formulation would provide a route toward complex response where membrane deformation involves multiple stable states or history-dependent mechanics. Finally, two-phase extensions would include capillary effects and air invasion, thereby opening toward enhanced prediction of flow redistribution not only related to xylem, but also in applied fields such as microfluidics and filtration devices~\cite{Beebe2002ARBE,Stone2004ARFM,Novak2020NatBiomedEng,Nagy2015MicrochemJ}.

\textbf{Acknowledgments}
P.G.L. acknowledges the University of Cagliari for financial support. G.A.Z. acknowledges the Italian Ministry of the University and Research via the ‘Rita Levi Montalcini’ grant (D.M. n. 1317, 15/12/2021, published on GU Serie Generale n. 226, 27/09/2022).
P.G.L. also thanks M.G. Badas for providing the computational resources needed to carry out the simulations. 

P.G.L. conceived the project. P.G.L. and G.A.Z. developed the theory and implemented the numerical methods. G.F. prepared the full-scale geometries and performed the microscopic, mesoscopic and vessel simulations. P.G.L. prepared and performed the remaining numerical simulations and the network-scale analysis. All authors discussed the results. P.G.L. wrote the draft. All authors revised the manuscript.

\textbf{Data Availability}.
The data that support the findings of this article are not publicly available. The data are available from the authors upon reasonable request.

\bibliographystyle{unsrt}
\bibliography{apssamp}

\end{document}